\documentclass[12pt,a4paper]{article}
\usepackage{amsmath}
\usepackage{amssymb}
\usepackage{slashed}
\usepackage{bbold}
\usepackage{wasysym}
\usepackage{graphicx}
\usepackage[usenames,dvipsnames]{color}
\usepackage[abs]{overpic}
\newcommand{\eqn}{equation}

\newcommand{\wt}{\widetilde}

\newcommand{\lb}{\left(}
\newcommand{\rb}{\right)}

\newcommand{\mO}{\mathcal{O}}
\newcommand{\ora}{\overrightarrow}

\newcommand{\GeV}{{\ensuremath\rm GeV}}
\newcommand{\TeV}{{\ensuremath\rm TeV}}
\newcommand{\pb}{{\ensuremath\rm pb}}
\newcommand{\fb}{{\ensuremath\rm fb}}

\newcommand{\ttt}[1]{\texttt{#1}}
\newcommand{\pT}{p_{\perp}}
\newcommand{\iteme}[1]{\item[\texttt{#1}\hfill]}
\newenvironment{entry}%
{\begin{list}{}{\setlength{\topsep}{0mm} \setlength{\itemsep}{0mm}
\setlength{\parskip}{0mm} \setlength{\parsep}{0mm}
\setlength{\leftmargin}{16mm} \setlength{\rightmargin}{0mm}
\setlength{\labelwidth}{14mm} \setlength{\labelsep}{2mm}}}%
{\end{list}}
\newenvironment{subentry}%
{\begin{list}{}{\setlength{\topsep}{0mm} \setlength{\itemsep}{0mm}
\setlength{\parskip}{0mm} \setlength{\parsep}{0mm}
\setlength{\leftmargin}{8mm} \setlength{\rightmargin}{0mm}
\setlength{\labelwidth}{14mm} \setlength{\labelsep}{2mm}}}%
{\end{list}}
\newcommand{\alphas}{\alpha_{\mathrm{s}}}
\newcommand{\kT}{k_{\perp}}
\newcommand{\q}{\mathrm{q}}
\newcommand{\qbar}{\overline{\mathrm{q}}}

\newcommand{\pTmin}{p_{\perp\mathrm{min}}}
\renewcommand{\d}{\mathrm{d}}
\newcommand{\itemc}[1]{\item[\textbf{#1}\hfill]}

\newcommand{\e}{\mathrm{e}}

\newcommand{\ee}{\e^+\e^-}
\newcommand{\pTmax}{p_{\perp\mathrm{max}}}
\newcommand{\p}{\mathrm{p}}
\begin{document}
\bibliographystyle{unsrt}
\thispagestyle{empty}
\def\thefootnote{\fnsymbol{footnote}}
\setcounter{footnote}{1}
\null
\hfill 
\vskip 0cm

\begin{center}

{\Large \boldmath{\bf $\sqrt{\hat{s}}_\text{min}$ resurrected
    }
    \par} \vskip 2.5em {\large
{\sc Tania Robens}\\[1ex]
{\normalsize \it SUPA, School of Physics and Astronomy,\\
University of Glasgow, Glasgow, G12 8QQ,  Scotland, UK \vspace{2mm}\\ and \vspace{2mm} \\ IKTP, TU Dresden,\\ Zellescher Weg 19, 01069 Dresden, Germany}}
\par \vskip 2em
\end{center}\par

\noindent{\bf Abstract:}\\[0.25em]
\noindent We discuss the use of the variable $\sqrt{\hat{s}}_\text{min}$, which
has been proposed in order to measure the hard scale of a multi parton
final state event using inclusive quantities only, on a SUSY data
sample for a $14\,\TeV$ LHC. In its original version, where this variable
was proposed on calorimeter level, the direct correlation to the hard
scattering scale
does not survive when effects from soft physics 
%in form of initial state radiation
%and underlying event 
are taken into account. We here show that 
when
using reconstructed objects instead of calorimeter energy and momenta
as input, 
%with a
%slight variation from its original definition,
%which is motivated by a
%later RECO level variation of  $\sqrt{\hat{s}}_\text{min}$, 
we manage to actually
recover this correlation for the parameter point considered here. 
%,which mainly consist in a direct link between the peak position of
%its distribution and the threshold of the hard matrix element. 
We furthermore discuss the effect of including $W\,+$ jets and $t\bar{t}\,+$ jets background in our analysis and the use of  $\sqrt{\hat{s}}_\text{min}$ for the suppression of SM induced background in new physics searches. %We discuss an analytic derivation of the peak position for a simple example.
\par
\null
\setcounter{page}{0}
\clearpage
\def\thefootnote{\arabic{footnote}}
\setcounter{footnote}{0}

\clearpage
\def\thefootnote{\arabic{footnote}}
\setcounter{footnote}{0}
\newpage
%\maketitle
%\section{Introduction}
\section{Introduction and motivation}
Since the startup and the following successful data taking of the
LHC, the LHC experiments have already published a large number of
result for exclusion limits for BSM physics, where for many BSM
scenarios the actual limits within specific parameter regions have
been strongly pushed to higher scales \cite{searches}. However, most
of these analyses have been performed within specific models, and more
generic variables, which provide information about the generic scale
of new physics without additional assumptions about the decay
topologies or decay chains, have not been fully
exploited. Furthermore, many variables which are currently proposed
for mass or scale determination for new physics processes only make
use of the transverse momentum of the event, thereby neglecting the
information which can be obtained by additionally including the
longitudinal information\footnote{Excellent reviews about different
  mass determination variables and their use, including advantages and
  disadvantages, have recently been published in
  \cite{Barr:2010zj,Barr:2011xt}.}. Examples for variables which make
use of transverse momentum only are eg
$M_{T2}$\cite{Lester:1999tx} or $M_{CT}$ \cite{Tovey:2008ui}. On the other hand, some more
traditional variables as eg invarant masses of composite objects as
used in edges studies \cite{Hinchliffe:1996iu,Bachacou:1999zb,:1999fr,Allanach:2000kt} implicitely use all visible information,
including the longitudinal momentum of the visible decay products. In \cite{Konar:2008ei} a new fully
inclusive calorimeter-level variable, $\sqrt{\hat{s}}_\text{min}$, was
proposed which promises to give information about the hard scale of
the underlying new physics processes without further assumptions or
specification of the decay products, and additionally makes use of the
longitudinal information of the inclusive event. In that paper, the
authors propose a conjecture which relates the peak position of
$\sqrt{\hat{s}}_\text{min}$ to the actual rise of the hard new physics
production cross section. However, this variable was subsequently
shown to have a strong dependence on the soft physics in terms of ISR
and underlying event. This is mainly caused by the fact that in this case,
the energy of the additional soft particles equally enters in the
calorimeter-level definition of $\sqrt{\hat{s}}_\text{min}$, which then boosts the
variable and its peak position to higher values with respect to the
parton-level quantity. In \cite{Konar:2008ei}, the authors tried to
circumvent this problem by introducing a pseudorapidity cut in order
to suppress the unwanted effects originating from soft physics from
entering  $\sqrt{\hat{s}}_\text{min}$. However, the introduction
of the cut destroyed the correlation between peak position and hard
cross section threshold which holds at parton level; this has explicitely been shown analytically for effects arising from initial state radiation (ISR) \cite{Papaefstathiou:2009hp, Papaefstathiou:2010ru}. In this case, the peak position is basically determined by the value of the pseudorapidity cut. Subsequently, $\sqrt{\hat{s}}_\text{min}$ was promoted to $\sqrt{\hat{s}}_\text{min}^\text{(reco)}$ \cite{Konar:2010ma}, using reconstructed objects at analysis level; for this variable, the correlation between its peak and the threshold of the hard production cross section was recovered, such that a determination of the hard scale of the BSM process was again made possible using experimentally accessible detector level objects.\\

Apart from providing information about the hard scale of the underlying parton level BSM process, new variables can equally be used as cut parameters for SM background suppression, and several of these variables have already made their way into the current BSM searches at the LHC experiments. It is therefore equally important to determine the use of $\sqrt{\hat{s}}_\text{min}$ for SM background suppression. Although this constitutes a slightly weaker use of the variable per se, it is still an important issue to investigate, especially as it has been proposed on a fully inclusive level and can therefore be applied without any further assumptions on the model or the specific decay chains and topology.\\

In this report, we therefore investigate the properties of  $\sqrt{\hat{s}}_\text{min}$ at analysis level using reconstructed objects. For this, we use a full sample for the mSugra point SPS1a\footnote{We are aware that this parameter point has recently been excluded by ATLAS measurements in the quark/ gluon plus missing transverse energy channel \cite{daCosta:2011qk}. However, we here want to show that the parton level  $\sqrt{\hat{s}}_\text{min}$ can actually be recovered with sufficient accuracy from analysis level objects. Spectra which evade current exclusion limits typically exhibit higher initial cascade particle masses, and our arguments are generically not affected by the actual position of the particle production threshold. This is only important in the studies of background suppression presented in Section \ref{sec:smbkg}; here, a higher peak value for the BSM induced variable should actually even enhance the SM background suppression which can be obtained using $\sqrt{\hat{s}}_\text{min}$.} \cite{Allanach:2002nj}, which contains all strong production as well as all decay chains; for this parameter point, our sample therefore corresponds to the full data set which would be obtained from strongly interacting initial cascade particles in a realization of SPS1a\footnote{Gaugino-gaugino initial cascade states, which have not been considered here, would contribute an additional $5\%$ to the total production cross section.}. We include soft physics in terms of initial and final state radiation (ISR/ FSR), as well as a fast detector simulation. We test the correlation between the threshold of the hard process and the peak of  $\sqrt{\hat{s}}_\text{min}$ on parton level for the inclusive sample as well as exclusive dominant final states. In addition, we show that, for our sample, this relation can be regained on reconstruction level using quite simple analysis object definitions, and that major discrepancies between analysis and parton level quantities can be traced back to uncertainties in the reconstruction of tau jets. We equally comment on the power of the analysis level variable for SM background suppression, and compare to similar variables. All analyses are done for a LHC-like proton proton collider with a center of mass energy of 14 \TeV~ and an integrated luminosity $\int\,\mathcal{L}\,=\,1\,\fb^{-1}$.\\

The report is organized as follows: in Section 2, we briefly review the variable definition of $\sqrt{\hat{s}}_\text{min}$ and define other kinematic quantities which were used in our study. In Section 3, we describe the data set we use in this study. Section 4 contains the comparison of parton and analysis level $\sqrt{\hat{s}}_\text{min}$, and Section 5 describes the inclusion of SM background and the use of  $\sqrt{\hat{s}}_\text{min}$ for SM background suppression, as well as a brief comparison with other (transverse) variables. We conclude in Section 6. 

\section{Variable definition}\label{sec:var}% and alternatives}
%\subsection{$\sqrt{\hat{s}}_\text{min}$ and RECO level $\sqrt{\hat{s}}_\text{min}$}
In this section, we will briefly review the original variable definition as well as its RECO level version; the interested reader is referred to \cite{Konar:2008ei, Konar:2010ma} for a more detailed discussion.\\
%\subsection{Use of $\sqrt{\hat{s}}_\text{min}$ on inclusive reconstruction level}

In general, $\sqrt{\hat{s}}_\text{min}$ is defined on an event by event basis as the minimal value for the partonic center-of-mass energy $\sqrt{\hat{s}}$ which is in agreement with the events' momentum configuration. It can be derived through a minimization process \cite{Konar:2008ei} as
\begin{\eqn}\label{eq:mastereq}
\sqrt{\hat{s}}_\text{min}\,\lb M_\text{inv} \rb\,=\,\sqrt{\slashed{E}_T^2\,+\,M_\text{vis}^2}\,+\,\sqrt{\slashed{E}_T^2\,+\,M^2_\text{inv}}
\end{\eqn}
with
\begin{\eqn*}
 M_\text{vis}^2\,=\,E^2-\overrightarrow{P}^2
\end{\eqn*}
 being the effective visible
mass and the invisible total mass 
\begin{\eqn}\label{eq:minv}
M_\text{inv}\,=\,\sum_{\text{invisible}}\,m
\end{\eqn}
 the sum over
all masses of invisible particles. For completeness, we give the minimization procedure leading to Eqn. (\ref{eq:mastereq}) in Appendix \ref{sec:mini}. Note that $M_\text{vis}$ and
  $M_\text{inv}$ are {\sl not} defined equivalently, and especially that $M_\text{inv}$ is {\sl not} the Lorentz-invariant mass of the total invisible system, but rather the sum over all invisible particles rest masses\footnote{For consistency, we here adopted the notation introduced in \cite{Konar:2008ei} for $M_\text{inv}$  and hope that the potentially misleading nomenclature does not cause confusion in the remainder of our discussion.}. In the
definition of $\sqrt{\hat{s}}_\text{min}$, $M_\text{inv}$ is therefore
an external input parameter, as it is the only quantity which cannot
be measured directly from experiment. Therefore, all results which are
derived in the following sections have an implicit dependence on the
value of $M_\text{inv}$. Throughout our study, we have usually set
this to its "true" BSM value
$M_\text{inv}\,=\,2\,m_{\wt{\chi}^0_1}\,(=\,193.4\,\GeV)$.\\ 

 The translation to experimentally accessible quantities is then straightforward and gives
\begin{\eqn}\label{eq:sqrtshat_meas}
\sqrt{\hat{s}}_\text{min}\lb M_\text{inv}\rb\,=\,\sqrt{E^2-P_Z^2}\,+\,\sqrt{\slashed{E}^{2}_T\,+\,M^2_\text{inv}}.
\end{\eqn}
Note that the use of transverse energy and momentum strongly depends on the definition of the specific quantity; we define
\begin{\eqn}\label{eq:ptslash}
\slashed{\ora{P}}_T\,=\,-\ora{P}_{T},\;\slashed{E}_{T}\,=\,|\slashed{\ora{P}}_T|
\end{\eqn}
where $E, P$ are the total energy and four momentum of all visible objects
\begin{\eqn}\label{eq:pvis}
P^\mu\,=\,\sum_\text{vis}p^\mu_{i},\;E\,=\,P^0,\,\ora{P}_T\,=\,\lb\begin{array}{c}P_X\\P_Y\end{array}\rb
\end{\eqn}
and the $z$-direction defines the beam-line.
 In the original proposal, all visible quantities are taken from calorimeters, and soft background is suppressed by a cut in the pseudorapidity $\eta$. Subsequently, for the {\sl correct} value of $M_\text{inv, true}$, a conjecture
\begin{\eqn}\label{eq:conj}
\left[\sqrt{\hat{s}}_\text{min} \lb M_\text{inv, true} \rb\,\right]_\text{peak}\sim\,\sqrt{s}_\text{th}
\end{\eqn}
is empirically derived, which links the peak position of $\sqrt{\hat{s}}_\text{min}$ to the actual threshold  $\sqrt{s}_\text{th}$ of the hard matrix element process. However, in \cite{Papaefstathiou:2009hp} it was subsequently pointed out that, using different values for the $\eta$ cut, the peak position for the calorimeter-based variable $\sqrt{\hat{s}}_\text{min}$ could actually be arbitrarily shifted around; the same effect has been observed in \cite{Brooijmans:2010tn}, which applies the original calorimeter-based definition of $\sqrt{\hat{s}}_\text{min}$ on our data sample. In answer to this criticism, new reconstruction and subsystem level variables were proposed in \cite{Konar:2010ma}. 
%These variables actually live up to the current (as well as past) status of collider physics analyses, where suppression techniques for soft physics from ISR, FSR and underlying event are applied prior to analyses\footnote{However, in applying $\sqrt{\hat{s}}_\text{min}$ for soft background suppression, one would of course make use of the original calorimeter based definition.}.  
%\\ 
We here use $\sqrt{\hat{s}}_\text{min}$ on an inclusive level
using reconstructed objects, which basically corresponds to the RECO
variable definition given in \cite{Konar:2010ma}. 
%We do not seek for a specific signature of the sample, ie we use all events which are present on a reconstruction level; on the other hand, we (hope to) sufficiently suppress the effects of soft physics by using reconstruction level events only. In this study, 
In our work, the suppression of effects from the parton shower has been achieved by quite simple
object-level definitions given in Table \ref{tab:objects}. 
%Differences
%between our results and the results presented in \cite{Konar:2010ma}
%are then mainly due to the difference in object definitions between
%the fast detector simulations which were used, as well as the fact
%that we applied additional constraints to our reconstruction level
%object definitions, cf. Table \ref{tab:objects}, which results in a more
%accurate reconstruction of the parton
%level quantities. 
We will show that we obtain the
parton-level  $\sqrt{\hat{s}}_\text{min}$ quite accurately in our
sample, and that we indeed observe a similar peak of the RECO-level
$\sqrt{\hat{s}}_\text{min}$ close to the production threshold for the
correct input value of $M_\text{inv}$. In addition, we investigate the actual sources of discrepancies between the parton
and reconstruction level $\sqrt{\hat{s}}_\text{min}$ in more
detail. In our work, we equally present the first study of
$\sqrt{\hat{s}}_\text{min}$ as a variable for SM background
suppression in BSM searches.\\

In this study, we use the
term ``leptons'' for all three SM lepton generations; in cases when we are concerned
with tau leptons alone we will mention this explicitly. Equally, the tau jets at analysis/ reconstruction level are defined by the tau jet reconstruction algorithm in Delphes \cite{Ovyn:2009tx} and differ from the {\sl parton level} tau lepton by the four-momenta of the invisible tau decay products, specifically the associated third generation neutrino. In the following, we use the term "tau" for the parton level and "tau jet" for the reconstruction level quantity, which for an ideal reconstruction of the visible tau decay products four-momenta only differ by the four-momentum of the invisible decay products.\\%, which
%we show to work for both correct and incorrect input values for $M_\text{inv}$.
\section{Data sample and event generation}\label{sec:data}
In this report, we have made use of the BSM data samples which have been generated in the course of the 2009 BSM Les Houches mass determination study; first results using these data for studies of various mass determination methods were presented in \cite{Brooijmans:2010tn}. We use a SUSY spectrum for the point SPS1a, where the spectrum was generated with the spectrum generator SOFTSUSY \cite{Allanach:2001kg}. Parton level events have been generated using Madgraph \cite{Stelzer:1994ta,Maltoni:2002qb} for the generation of the heavy initial cascade particles (i.e. the squark-squark, squark-gluino, and gluino-gluino initial states). The heavy pair-produced particles have then been fully decayed according to the respective branching ratios into all possible decay products using Bridge \cite{Meade:2007js} within the Madgraph framework; we therefore consider a complete sample for this parameter point, which contains {\sl all} possible final states. The SM background has been generated using Alpgen \cite{Mangano:2002ea}. For parton shower and hadronization, we used Pythia \cite{Sjostrand:2006za}, where the parton shower evolution follows the Pythia 6.4 default, ie is $Q^2$ ordered with additional modifications to guarantee color coherence, as well as matrix element corrections where these are available (Pythia switches are given in Appendix \ref{app:pshower}). The detector simulation has been performed with Delphes \cite{Ovyn:2009tx} in its default mode. For specific input parameters and setups, we refer to the specifications which can be found in the data base for our samples \cite{samples}. Data analysis as well has fitting has been done within the ROOT \cite{Antcheva:2009zz,root} framework.
All our results have been obtained with a data sample for a
center-of-mass energy $\sqrt{S}_\text{hadr}\,=\,14\,\TeV$ and an
integrated luminosity $\int\mathcal{L}\,=\,1\,\fb^{-1}$. Table
\ref{tab:xsecs} lists the production cross sections for the hard
$2\,\rightarrow\,2$ process; these numbers were obtained using the
Madgraph parton level $2\,\rightarrow\,2$ production cross sections,
with the electroweak scale spectrum obtained from SOFTSUSY, convoluted
with PDFs to account for the parton to hadron transition for the
incoming states. For this study, we restrict ourselves to the leading order predictions for the hard process in both signal and background simulation\footnote{A fully differential study including NLO contributions to account for cut effects would require $2\,\rightarrow\,n$ event generators for both BSM signal and SM background, which additionally include the matching of parton shower and NLO contribution; although fast progress has been made in this field for SM processes, no fully differential higher order BSM generator is currently publicly available.}.
%\footnote{To our knowlegde, no $2\,\rightarrow\,n$ parton level NLO generator is available which can handle BSM processes; even for SM predictions, a combination of NLO predictions and parton showers, which is required for a full treatment of differential distributions, is only available for selected processes (cf eg \cite{}).}.
\begin{table}
\begin{\eqn*}
\begin{array}{l|c}%|c}
X_{1}X_{2}&2\,\rightarrow\,2\\ \hline%&2\,\rightarrow\,3\\ \hline
\tilde{q}\tilde{q}&6.56\\%&7.83\\
\tilde{q}\tilde{g}&19.96\\%&21.75\\
\tilde{g}\tilde{g}&4.53\\%&5.47\\
%\wt{\chi}\wt{\chi}\,(j)&1.97% &4.89
\end{array}
\end{\eqn*}
\caption{\label{tab:xsecs} SPS1a production cross sections in $\pb$ for $p\,p\,\rightarrow\,X_{1}\,X_{2}$ using Madgraph $2\,\rightarrow\,2$ parton level production cross sections, convoluted with PDFs, for a hadronic center-of-mass energy of $14\,\TeV$. CTEQ6L1 PDFs \cite{Stump:2003yu} were used. 
%$2\,\rightarrow\,3$ sample includes an explicitly generated hard jet, where hard is defined by $p_{T,\rm{jet}}\,>\,40\,\GeV$. 
}
\end{table} 
\section{Parton and analysis level $\sqrt{\hat{s}}_\text{min}$}\label{sec:pandana}
As already discussed in Section \ref{sec:var}, the variable $\sqrt{\hat{s}}_\text{min}$ has undergone several developments since its original proposal. Initially defined as a calorimeter-based variable, it was shown to be quite sensitive to effects of soft physics for the respective processes. Especially the original merit of this variable, namely the correlation of the peak position and the threshold of the heavy pair-produced particles at the beginning of the decay chain, is strongly influenced by the soft physics of the event. The original suggestion of the authors was to introduce a cut on the pseudorapidity; however, the authors in \cite{Papaefstathiou:2009hp,Papaefstathiou:2010ru} have shown analytically that the position of the peak position in this case is completely cut-value dependent; similar results have been observed in \cite{Brooijmans:2010tn}.\\

In this study, we show that, if $\sqrt{\hat{s}}_\text{min}$ is defined at analysis object level rather than on calorimeter level, the parton level variable $\sqrt{\hat{s}}_\text{min}$ can be reconstructed quite well using simple object definitions. This recovery of the parton level peak position using a reco-level variable for both inclusive and exclusive final states, however for different parameter points, have equally been presented in \cite{Konar:2010ma}. 
%In our study, 
%by using more stringent requirements for the analysis level object definitions,
%we not only manage to recover the peak position, but equally do not obtain a long distribution tail at higher $\sqrt{\hat{s}}_\text{min}$ values. 
For our sample, we equally observe that the conjectured correlation between the rise of the hard scattering event cross section and the peak position of  $\sqrt{\hat{s}}_\text{min}$ holds; however, we want to emphasize that this is on the level of a {\sl conjecture} which has not been systematically studied or proven on an analytic level, although some preliminary studies indicate a kinematic origin which emerges after the convolution with PDFs\footnote{This result has been obtained with a unit matrix element as well as unit PDFs; in this case, the peak position of $\sqrt{\hat{s}}_\text{min}$ arises from the lower PDF integration boundary following from the kinematic lower limit which guarantees that $\sqrt{s}_\text{part}\,\geq\,\sqrt{s}_\text{threshold}$. More realistic scenarios with non-uniform PDFs and matrix elements are currently under investigation. } \cite{meprep}. Therefore, even on parton level, it is currently  unclear whether this conjecture necessarily holds for all BSM parameter points and scenarios.
%\footnote{Indeed it has recently been shown \cite{Conley:2010du} that a similar conjecture between $M_\text{eff}$ and the hard parton cross section threshold does not automatically hold; a similar study of $\sqrt{\hat{s}}_\text{min}$ is still lacking.}. 
Equally, this conjecture only holds for a correct input value of $M_\text{inv}$.\\  

\begin{table}
\begin{center}
{\small
\begin{tabular}{c|c|c}
{\sl object}&{\sl Delphes predefinition}&{\sl additional requirement}\\
\hline \hline
electron/ position& $|\eta|\,<\,2.5$ in tracker, $p_T\,>\,10\,\GeV$&isolated\\ %\hline
%&&\\
muon& $|\eta|\,<\,2.4$ in tracker, $p_T\,>\,10\,\GeV$&isolated\\ %\hline
%&&\\
lepton isolation criteria& no track with
$p_T > 2\, \GeV$ &no track with
$p_T > 6\, \GeV$\\
&in a cone with $dR=0.5$&in a cone with  $dR=0.5$\\
&around the considered lepton&around the considered lepton\\ %\hline
%&&\\
$n$ leptons&-----& exactly $n$ isolated leptons\\
&& at detector level\\
taujet& $p_T\,>\,10\,\GeV$&-----\\ %\hline
%&&\\
jet& $p_T\,>\,20\,\GeV$&$p_{T,\rm{jet}}\,>\,50\,\GeV$, $|\eta|_{\rm{jet}}\,<\,3$  \\
&CDF jet cluster algorithm \cite{Abe:1991ui}, &\\
&$R\,=\,0.7$&\\ %\hline
%&&\\
 Missing transverse&-----&$E_T^{\rm{miss}}\,>\,100\,\GeV$\\
 energy&&
\end{tabular}}
\caption{\label{tab:objects} Physical object definitions in terms of the single objects pseudorapidity $\eta$, absolute value of transverse momentum $p_T$, and (jet) cone radius $R$ for analysis level objects on detector level. We basically adapt the Delphes predefinitions, with slightly more stringent requirements for isolated leptons and jet definitions. We equally set a lower limit ${E}_T^{\text{miss}}\,\geq\,100\,\GeV$ for events with missing transverse energy.}
\end{center}
\end{table}
In the following, we will compare quantities derived on the parton level with the same quantities which have been derived from analysis level objects. For the identification of the former, we consider the hard process, i.e. our data sample after the complete decay to SM particles and the LSP, but before the parton shower, hadronization, and detector simulation. All particles are considered as visible apart from neutrinos and the LSP. The invisible total four-momentum is then the sum of the latter particles' four-vectors
\begin{\eqn*}
P_\text{invis}^\text{parton}\,=\,\sum_{\nu's,\wt{\chi}^{0}_{1}}p_{i},
\end{\eqn*}
and the same holds for the missing transverse momentum. At analysis level,
we require all physical objects to fulfill the object definition requirements given in Table \ref{tab:objects} on detector level; these object definitions basically follow the Delphes predefinitions, where we introduced slightly more stringent requirements for lepton isolation and jet criteria and equally set a lower limit of $100\,\GeV$ on the total missing energy\footnote{These cuts closely follow cuts used in the SUSY analysis studies in \cite{:1999fr,Aad:2009wy}. Due to the relatively high $p_T$ jet cuts, together with a high $E^\text{miss}_T$ cut and lepton isolation criteria, we expect minimum bias events to be sufficiently suppressed (\cite{privcomm}, as well as section 6.1 in \cite{Moraes:2007rq}). }. Visible and invisible quantities are then defined according to Eqns. (\ref{eq:ptslash}) and (\ref{eq:pvis}) in Section \ref{sec:var}.\\

\begin{figure} 
\begin{center}
%\put(50,50){$a$}
\begin{overpic}[width=0.45\textwidth, angle=-90]{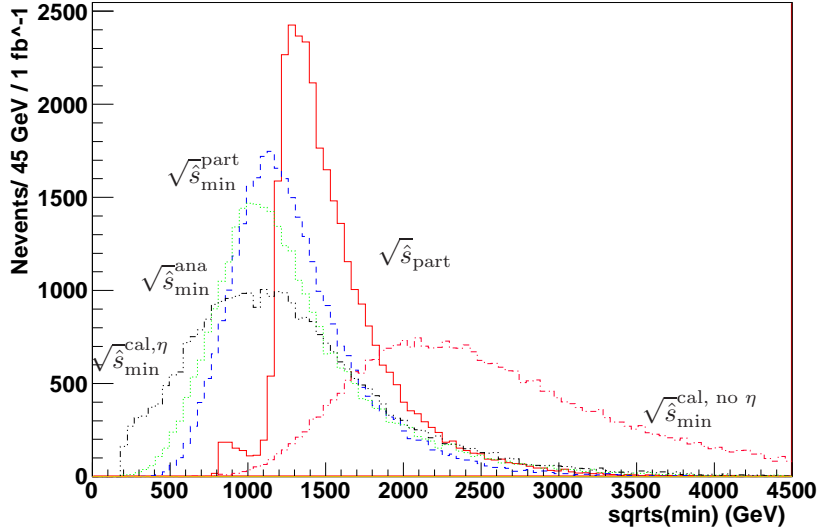}
%\put(0,100){huhu}
%\put(100,100){huhu2}
\put(135,100){{ \footnotesize $\textstyle \sqrt{\hat{s}}_\text{part}$}}
\put(55,130){{ \footnotesize $\textstyle \sqrt{\hat{s}}_\text{min}^\text{part}$}}
\put(45,90){{ \footnotesize $\textstyle \sqrt{\hat{s}}_\text{min}^\text{ana}$}}
\put(235,40){{ \footnotesize  $\textstyle \sqrt{\hat{s}}_\text{min}^{\text{cal, no } \eta}$ }}
\put(30,60){{\footnotesize $\textstyle \sqrt{\hat{s}}_\text{min}^\text{cal,$\eta$}$}}
%\put(0,160){\includegraphics[angle=-90, width=0.6\textwidth]{plots/true_all_all.eps}}
\end{overpic}
\end{center}
\caption{\label{fig:allcol_noreq} Sum of $\tilde{q}\tilde{q},\,\tilde{q}\tilde{g},\,$, and $\tilde{g}\tilde{g}$ initial states, where $\tilde{q}\tilde{g}$ initial states dominate. All final states which fulfill object definitions from Table \ref{tab:objects} are included. True $\sqrt{\hat{s}}$ ({red}; solid), parton level $\sqrt{\hat{s}}_\text{min}$ ({blue}; dashed), analysis level $\sqrt{\hat{s}}_\text{min}$ ({green}; dotted), $\sqrt{s}_\text{min}$ using calorimeters ({pink}; dash-dotted) and same with an $|\eta|<1.4$ cut (black; dash-dot-dot-dotted). $\sqrt{S}_\text{hadr}\,=\,14\,\TeV,\,\int\mathcal{L}\,=\,1\,\fb^{-1}$; corresponds to 31050 events. Shift between parton level and analysis level peak is about 70 \GeV. The calorimeter based distribution without a pseudorapidity cut exhibits a peak at much larger $\sqrt{\hat{s}}_\text{min}$ values.}
\end{figure}

We first study the variable $\sqrt{\hat{s}}_\text{min}$ for a complete inclusive sample, i.e. we sum over all final states of the hard process. Our main results are shown in Figure \ref{fig:allcol_noreq}, where we compare the true $\sqrt{\hat{s}}$, parton level $\sqrt{\hat{s}}_\text{min}$, reconstruction level $\sqrt{\hat{s}}_\text{min}$ as well as the original calorimeter based variable with and without a cut in pseudo rapidity $\eta$, as originally suggested in \cite{Konar:2008ei}. We see that the parton level $\sqrt{\hat{s}}_\text{min}$ peaks quite close to the actual heavy particle production threshold as suggested in \cite{Konar:2008ei}; equally, we observe that the same variable from reconstructed objects again peaks close to the threshold, but is shifted to slightly lower values with respect to the parton level quantity. We will comment on this in more detail below. In contrast, the pure calorimeter based variable exhibits a peak position at quite high values and can therefore not be used for a scale measurement of the new physics process. Restricting the contributions to calorimeter energy deposits with a minimal pseudorapidity improves this behavior and brings the peak closer to lower values; however, this approach suffers from the drawbacks pointed out in  \cite{Papaefstathiou:2009hp,Papaefstathiou:2010ru}. \\

\begin{figure} 
\begin{center}
\begin{overpic}[width=0.45\textwidth, angle=-90]{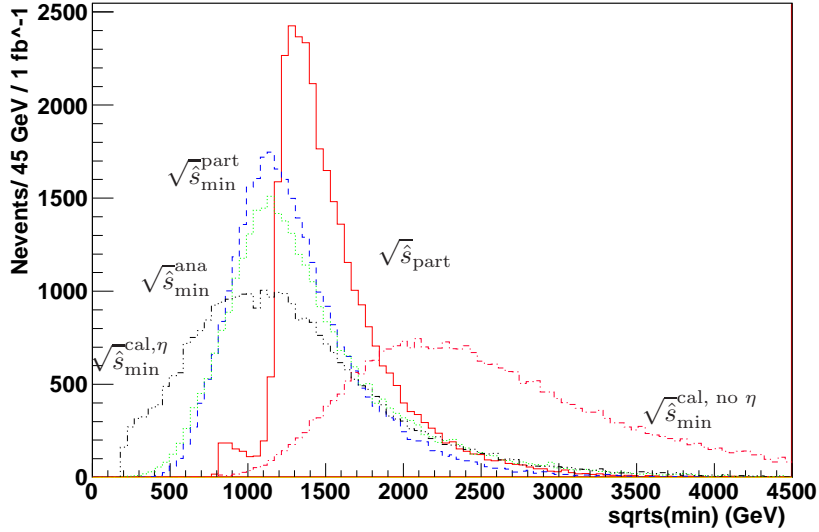}
%\put(0,100){huhu}
%\put(100,100){huhu2}
\put(135,100){{ \footnotesize $\textstyle \sqrt{\hat{s}}_\text{part}$}}
\put(55,130){{ \footnotesize $\textstyle \sqrt{\hat{s}}_\text{min}^\text{part}$}}
\put(45,90){{ \footnotesize $\textstyle \sqrt{\hat{s}}_\text{min}^\text{ana}$}}
\put(235,40){{ \footnotesize  $\textstyle \sqrt{\hat{s}}_\text{min}^{\text{cal, no } \eta}$ }}
\put(30,60){{\footnotesize $\textstyle \sqrt{\hat{s}}_\text{min}^\text{cal,$\eta$}$}}
%\put(0,160){\includegraphics[angle=-90, width=0.6\textwidth]{plots/true_all_all.eps}}
\end{overpic}
\end{center}
\caption{\label{fig:true_all_truetaus} As Fig.~\ref{fig:allcol_noreq}, where the hard matrix element tau four-vectors were used for the analysis level $\sqrt{\hat{s}}_\text{min}$. With the differences due to tau identification at the analysis level removed, parton and analysis level peak positions agree within error bars. Explicit numbers are given in Section \ref{sec:pandana}.}
\end{figure}

For a more accurate determination of the peak position and a viable assessment of the error in its position, we fit the  $\sqrt{\hat{s}}_\text{min}$ distribution with a Gaussian around its peak, where we use the largest fit region which is still in agreement with $\chi^{2}/ \text{d.o.f}\,\sim\,\mO(1)$. Specifically, we use $600\,\GeV\,\leq\,\sqrt{\hat{s}}_\text{min}\,\leq\,1400\,\GeV$ and $400\,\GeV\,\leq\,\sqrt{\hat{s}}_\text{min}\,\leq\,1400\,\GeV$ to determine the parton level and analysis level peak positions respectively. We then obtain
\begin{eqnarray*}
\text{parton level }\sqrt{\hat{s}}_\text{min}^\text{peak}&:&(1152\,\pm\,4)\,\GeV\\
\text{analysis level }\sqrt{\hat{s}}_\text{min}^\text{peak}&:&(1083\,\pm\,4)\GeV\\
\end{eqnarray*}
 We see that the reconstruction level variable for the overall sample peaks close to the "true" maximum of the parton level variable, the difference being $\mO(100\,\GeV)$. In order to pin down the major sources of this shift, we have performed detailed studies for specific final state signatures; we will discuss this in more detail in Section \ref{sec:signals}.
%\input{signals}
%\subsection{Result for completely inclusive sample}
To summarise the result of this section, we observe that, in our sample, larger shifts
in the peak positions stem from processes with one or more leptons in
the final state. 
%For these, this effect can indeed be traced back to
%differences in the total visible energy of the respective
%processes. 
One source of this is the imperfect reconstruction of tau jets from parton to analysis level objects. We
can test this by taking an "idealistic" approach, where we use the
parton level four-vector values for taus in the analysis level
objects; this simple "gedankenexperiment" trick, where we assume a
perfect reconstruction of tau jets at analysis level,
significantly reduces this difference, cf. Figure
\ref{fig:true_all_truetaus}. While such a requirement is in fact not
possible in reality, it however shows that our (quite loose) lepton
definitions and resulting poor tau reconstruction are a major source of this shift, and more dedicated algorithms might further reduce this discrepancy. Fitting the "new" analysis level distribution within the range $600\,\GeV\,\leq\,\sqrt{\hat{s}}_\text{min}\,\leq\,1400\,\GeV$, we obtain
\begin{eqnarray*}
\text{analysis level }\sqrt{\hat{s}}_\text{min}^\text{peak},\,\tau\,=\,\tau_\text{parton}&:&(1163\,\pm\,4)\,\GeV
\end{eqnarray*}
and we see that the discrepancy with the parton level value of
$(1152\,\pm\,4)\,\GeV$ reduces to the permill level, cf. Fig.~\ref{fig:true_all_truetaus}. The average heavy particle threshold in our sample is
\begin{eqnarray*}
\text{true (average) }\overline{(m_{1}+m_{2})}&:&1146\,\GeV\\
\end{eqnarray*}
which again agrees with the parton level value of
$\sqrt{\hat{s}}^\text{peak}_\text{min}$ on permill level within the error bars \footnote{In this work, we only want to demonstrate that the approximate peak position of the parton level variable can actually be obtained from analysis level objects; for more dedicated analyses, the peak position could also be determined by other means, eg. a fit to a more variable-specific function.}. In addition, the object definitions in Table \ref{tab:objects} equally allow for an adequate recovery of the parton level distribution shape, and, more specifically, we are able to suppress distribution tails for higher $\sqrt{\hat{s}}_\text{min}$ values appearing in the reco-level definition of this variable in \cite{Konar:2010ma}. A breakdown in terms of pairs of initially produced particles prior to the cascade decays is given in Table \ref{tab:inis}.
\begin{table}
\begin{\eqn*}
\begin{array}{c||c||c||c|c}
\text{initial cascade particles}&\text{threshold}&\sqrt{\hat{s}}_\text{min}^\text{peak;parton}&\sqrt{\hat{s}}_\text{min}^\text{peak;ana}&\sqrt{\hat{s}}_\text{min}^\text{peak;ana,$\tau\,=\,\tau_{p}$}\\ 
&[\GeV]&[\GeV]&[\GeV]&[\GeV]\\ \hline \hline
\tilde{g}\tilde{q}&1163&1195\,\pm\,5&1101\,\pm\,6&1204\,\pm\,6\\
&&1170&1035&1170\\ \hline
\tilde{q}\tilde{q}&1046&1042\,\pm\,5&1012\,\pm\,8&1088\,\pm\,8\\
&&1080&1035&1080\\ \hline
\tilde{g}\tilde{g}&1215&1257\,\pm\,7&1150\,\pm\,7&1241\,\pm\,6\\
&&1260&1170&1260
\end{array}
\end{\eqn*}
\caption{\label{tab:inis} Peak positions for separate heavy initial cascade particles for parton level ($\scriptstyle \sqrt{\hat{s}}_\text{min}^\text{peak;parton} $) and analysis level ($\scriptstyle \sqrt{\hat{s}}_\text{min}^\text{peak;ana}$)  quantities as well as analysis level quantity for idealized tau jets (${\scriptstyle \sqrt{\hat{s}}_\text{min}^\text{peak;ana,$\scriptstyle \tau\,=\,\tau_{p}$}}$) in \GeV, where the value in the respective first line arises from a Gaussian fit around the peak, while the second corresponds to the more simplified definition of the peak position by maximal number of bin entries. In addition, we give the average threshold value $\sqrt{s}_\text{th}\,=\,\overline{(m_1+m_2)}$ for each sample, where $m_{1,2}$ are the masses of the heavy initial cascade particles. We see that the peak position from both peak position definitions are close to the actual thresholds; in addition, the effect of imperfect tau reconstruction account for an approximate shift $\mO(100\,\GeV)$ for all initial state pairings.}
\end{table}
%\begin{figure} 
%\begin{center}
%\includegraphics[angle=-90, width=0.6\textwidth]{plots/true_2l3j.eps}
%\end{center}
%\caption{as Fig. \ref{fig:all2j2l} $\int\mathcal{L}\,=\,1\,\fb^{-1}$, exactly 2 leptons in the final state, 3 hardest jets (not on calorimeter level though). Corresponds to 2092 events.}
%\end{figure}
\\

\begin{figure} 
\begin{center}
\begin{overpic}[width=0.45\textwidth, angle=-90]{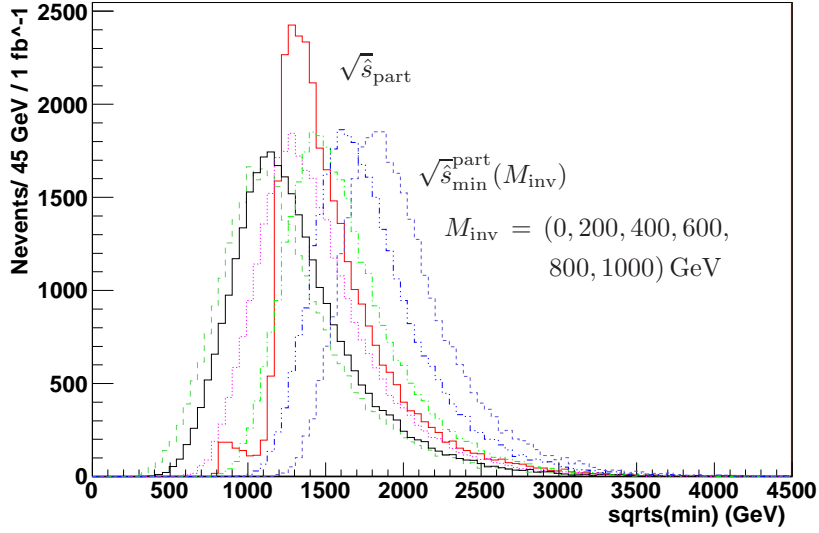}
%\put(0,100){huhu}
%\put(100,100){huhu2}
\put(120,170){{ \footnotesize $\textstyle \sqrt{\hat{s}}_\text{part}$}}
\put(150,130){{ \footnotesize $\textstyle \sqrt{\hat{s}}_\text{min}^\text{part}(M_\text{inv})$}}
\put(160,110){{ \footnotesize $M_\text{inv}\,=\,(0,200,400,600,$}}
\put(165,95){{ \footnotesize $\phantom{M_\text{inv}\,=\,}800,1000)\,\GeV$}}
%\put(0,160){\includegraphics[angle=-90, width=0.6\textwidth]{plots/true_all_all.eps}}
\end{overpic}
\end{center}
\caption{\label{fig:allparton} Parton level $\sqrt{\hat{s}}_\text{part}$ {\small (red; solid)}, and parton level $\sqrt{\hat{s}}_\text{min}^\text{part}$ dependence on the input value $M_\text{inv}$ for varying values. Shown are results for $M_\text{inv}\,=\,0\,\GeV\,\text{\small (dark green; long dashed)}$, $200\,\GeV\,\text{\small (black; solid)}$, $400\,\GeV\,\text{\small (pink; dotted)}$,   $600\,\GeV\,\text{\small (green; dash-dotted)}$, $800\,\GeV\,\text{\small (dark blue; dash-dot-dot-dotted)}$, $1000\,\GeV\,\text{\small (light blue; short dashed)}$. The corresponding peak positions for increasing $M_\text{inv}$ input values as given above, using a maximal bin definition, are obtained as $\sqrt{\hat{s}}_\text{min}^\text{peak; part}\,=\,(1035,\,1170,\,1305,\,1440,\,1620,\,1845)\,\GeV$. As before, the analysis level peak positions using the parton level tau leptons coincide with the parton level peaks (not shown here).}
\end{figure}

For illustration purposes and completeness, we also investigate the
parton level $\sqrt{\hat{s}}_\text{min}$ dependence on the input value
for $M_\text{inv}$; similar results have already been presented in
\cite{Konar:2008ei} and \cite{Brooijmans:2010tn} for the analysis
level quantity. For R-parity conserving SUSY scenarios, as considered
in this study,  this corresponds to the guess of the LSP mass, as in
this case $M_\text{inv}\,=\,2\,m_\text{LSP}$ when neutrino masses are neglected. Figure \ref{fig:allparton} shows the shift of the parton level $\sqrt{\hat{s}}_\text{min}$ distribution for different input values $M_\text{inv}$. We see that a variation of the $M_\text{inv}$ mass leads to a shift in the peak of a similar magnitude. Therefore, we again emphasize that the results presented in this study concerning the correlation of the peak of the $\sqrt{\hat{s}}_\text{min}$ distribution and the hard scale of the underlying production process have indeed an implicit dependence on the {\sl correctness} of the guessed input value for $M_\text{inv}$, as already discussed in the original proposal of this variable \cite{Konar:2008ei}, and therefore generically only allow for a measurement of the hard scale as a function of this variable\footnote{Several other widely used variables, as eg the original definition of $M_{T2}$ \cite{Lester:1999tx}, equally exhibit a dependence on an input value for the LSP mass.}. As before, the parton level distributions and peak positions could be reproduced using analysis level objects in the idealized version, ie replacing the analysis level tau-jets with parton-level tau leptons in $\sqrt{\hat{s}}_\text{min}^\text{ana}$.
 % only include if needed !!
\subsection{Signal based searches}\label{sec:signals}
In this section, we consider the variable $\sqrt{\hat{s}}_\text{min}$ for several exclusive final states. The (parton level) dominant decay modes of our sample are given in Table \ref{tab:domhard_decay}.
\begin{center}
\begin{table} 
\begin{\eqn*}
\begin{array}{c|c|c|c}
\text{final states, hard matrix element}&\text{main source}&N_\text{hard}&N_\text{ana}\\ \hline
\text{0 leptons, 3 jets}&\tilde{q}\tilde{g}&4480&14247\\
\text{2 leptons, 3 jets}&\tilde{q}\tilde{g} (97\%)&4020&2092\\
\text{1 lepton, 3 jets}&\tilde{q}\tilde{g} (99.99\%)&3740&5282\\
\text{2 leptons, 2 jets}&\tilde{q}\tilde{q}&1776&2745\\
\text{1 lepton, 2 jets}&\tilde{q}\tilde{q}&1366&6997
\end{array}
\end{\eqn*}
\caption{\label{tab:domhard_decay} Number of events for dominant parton level decay modes, characterized by specific visible final states, on parton level ($N_\text{hard}$) and at analysis level ($N_\text{ana}$). At analysis level, the jet number requirement for event selection is changed from an exact equality to a minimal number of jets. If not stated otherwise, the main source provides all events with a specific signature on the parton level. Examples for dominant decay chains leading to the specific parton-level final states are given in Section \ref{sec:signals}.}
\end{table}
\end{center}
Most of the dominant final states can be tracked down to a couple of competing processes, and can be broken down to the following parton-level decay chains
\begin{itemize}
\item{$\tilde{q}\tilde{g}$, 3 jet channel}\\
\begin{eqnarray*}
\tilde{q}_{R}\,\tilde{g}&\rightarrow&q_{R}\bar{q}_{R}\tilde{q}_{R}\,\wt{\chi}^{0}_{1}\,\rightarrow\,q_{R}\bar{q}_{R}\,q_{R}\,\wt{\chi}^{0}_{1}\wt{\chi}^{0}_{1}\;\;\;(90\%)
\end{eqnarray*}
\item{$\tilde{q}\tilde{g}$, 3 jet 2 lepton channel}\\
\begin{eqnarray*}
\tilde{q}_{L}\,\tilde{g}&\rightarrow&q_{L}\tau^{+}\tau^{-}\,\wt{\chi}^{0}_{1}\,q'_{R}\bar{q}'_{R}\wt{\chi}^{0}_{1}\;\;\;(27\%)\\
\tilde{q}_{R}\,\tilde{g}&\rightarrow&q_{R}\wt{\chi}^{0}_{1}\,b\,\bar{b}\tau^{+}\tau^{-}\,\wt{\chi}^{0}_{1}\;\;\;(22\%)\\
\tilde{q}_{R}\,\tilde{g}&\rightarrow&q_{R}\wt{\chi}^{0}_{1}\,q'_{L}\,\bar{q}'_{L} \tau^{+}\tau^{-}\,\wt{\chi}^{0}_{1}\;\;\;(17\%)\\
\tilde{q}_{L}\,\tilde{g}&\rightarrow&q'_{L}\,\tau\,\nu_\tau\wt{\chi}^{0}_{1}\,q''_{L}\,\bar{q}'''_{L} \tau\,\nu_\tau\wt{\chi}^{0}_{1}\;\;\;(17\%)\\
\tilde{q}_{L}\,\tilde{g}&\rightarrow&q'_{L}\,\tau\,\nu_\tau\wt{\chi}^{0}_{1}\,b\,\bar{t}\tau\,\nu_\tau\wt{\chi}^{0}_{1}\;\;\;(17\%)\\
\end{eqnarray*}
\item{$\tilde{q}\tilde{g}$, 3 jet 1 lepton channel}\\
\begin{eqnarray*}
\tilde{q}_{R}\,\tilde{g}&\rightarrow&q_{R}\wt{\chi}^{0}_{1}\,b\,\bar{t}\tau\nu_\tau\,\wt{\chi}^{0}_{1}\;\;\;(45\%, \text{BR}\,\sim\,0.09)\\
\tilde{q}_{R}\,\tilde{g}&\rightarrow&q_{R}\wt{\chi}^{0}_{1}\,q_{L}\,\bar{q}'_{L}\tau\nu_\tau\,\wt{\chi}^{0}_{1}\;\;\;(30\%, \text{BR}\,\sim\,0.06)\\
\tilde{q}_{L}\,\tilde{g}&\rightarrow&q'_{L}\,\tau\,\nu_\tau\wt{\chi}^{0}_{1}\,q''_{R}\,\bar{q}''_{R}\wt{\chi}^{0}_{1}\;\;\;(25\%, \text{BR}\,\sim\,0.05)\\
\end{eqnarray*}
\item{$\tilde{q}\tilde{q}$, 2 jet 2 lepton channel}\\
\begin{eqnarray*}
\tilde{q}_{L}\,\tilde{q}'_{L}&\rightarrow&q_{L}''\tau\nu_\tau\wt{\chi}^{0}_{1}\,q_{L}'''\,\tau\nu_\tau\,\wt{\chi}^{0}_{1}\;\;\;(36\%)\\
\tilde{q}_{R}\,\tilde{q}'_{L}&\rightarrow&q_{R}\wt{\chi}^{0}_{1}\,q'_L\,\tau^{+}\tau^{-}\wt{\chi}^{0}_{1}\;\;\;(64\%)\\
\end{eqnarray*}
\item{$\tilde{q}\tilde{q}$, 2 jet 1 lepton channel}\\
\begin{eqnarray*}
\tilde{q}_{R}\,\tilde{q}'_{L}&\rightarrow&q_{R}\wt{\chi}^{0}_{1}\,q''_L\,\tau\nu_\tau\wt{\chi}^{0}_{1}\;\;\;(100\%)\\
\end{eqnarray*}
\end{itemize}
At the analysis/ reconstruction level, we here require to have a {\sl minimal} jet multiplicity, which leads to much larger event numbers especially for signatures with a smaller number of leptons. We equally do not apply any dedicated additional channel-based cuts. \\

\begin{figure} 
\begin{minipage}{0.5\textwidth}
\begin{center}
\begin{overpic}[width=0.65\textwidth, angle=-90]{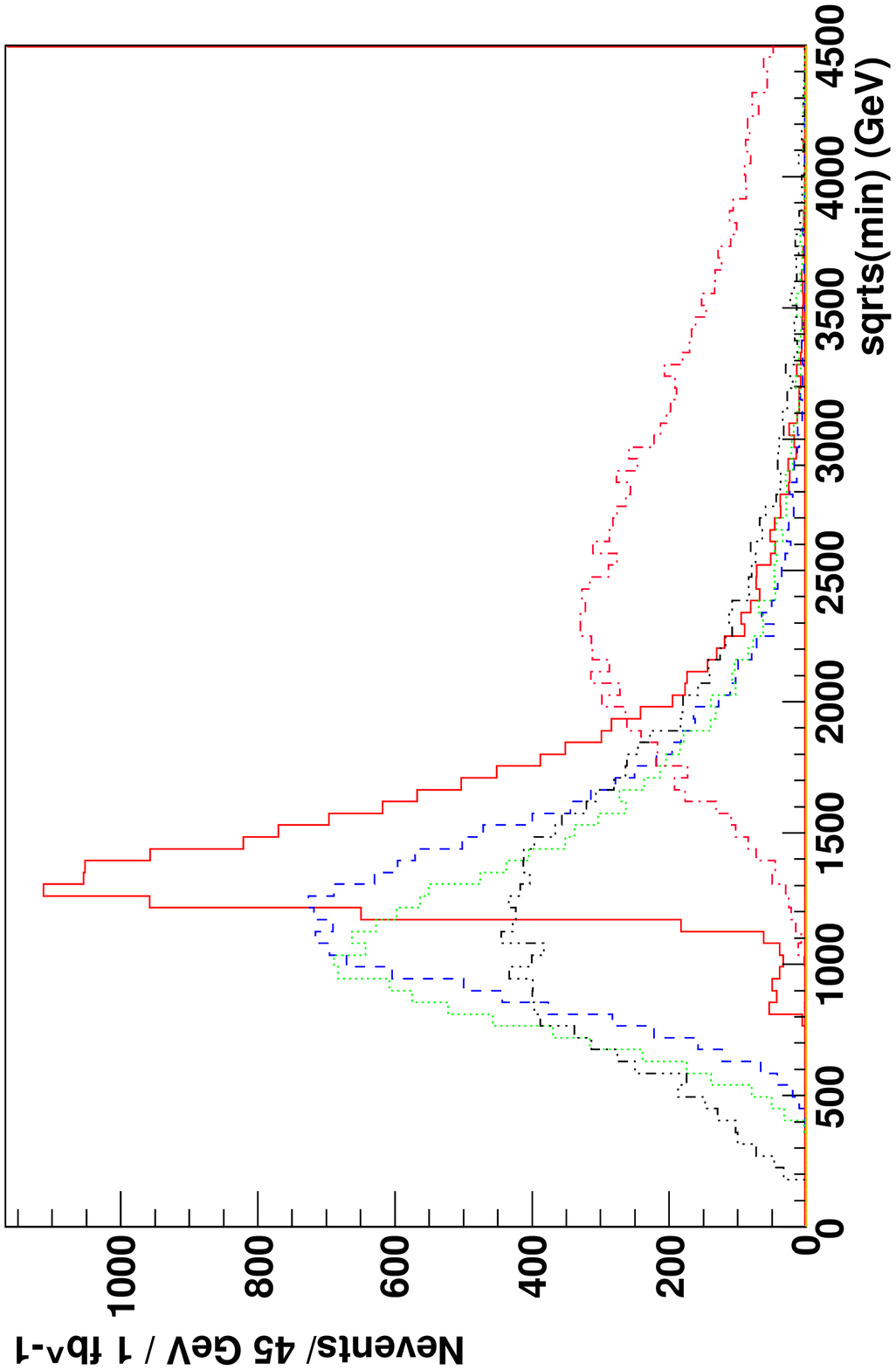}
%\put(0,100){huhu}
%\put(100,100){huhu2}
\put(90,95){{ \footnotesize $\scriptstyle \sqrt{\hat{s}}_\text{part}$}}
\put(40,100){{\footnotesize $\scriptstyle \sqrt{\hat{s}}_\text{min}^\text{part}$}}
\put(30,65){{\footnotesize $\scriptstyle \sqrt{\hat{s}}_\text{min}^\text{ana}$}}
\put(165,35){{\footnotesize  $\scriptstyle \sqrt{\hat{s}}_\text{min}^{\text{cal, no } \eta}$ }}
\put(20,40){{\footnotesize $\scriptstyle \sqrt{\hat{s}}_\text{min}^\text{cal,$\eta$}$}}
\end{overpic}
\end{center}
\end{minipage}
\begin{minipage}{0.5\textwidth}
\begin{center}
\begin{overpic}[width=0.65\textwidth, angle=-90]{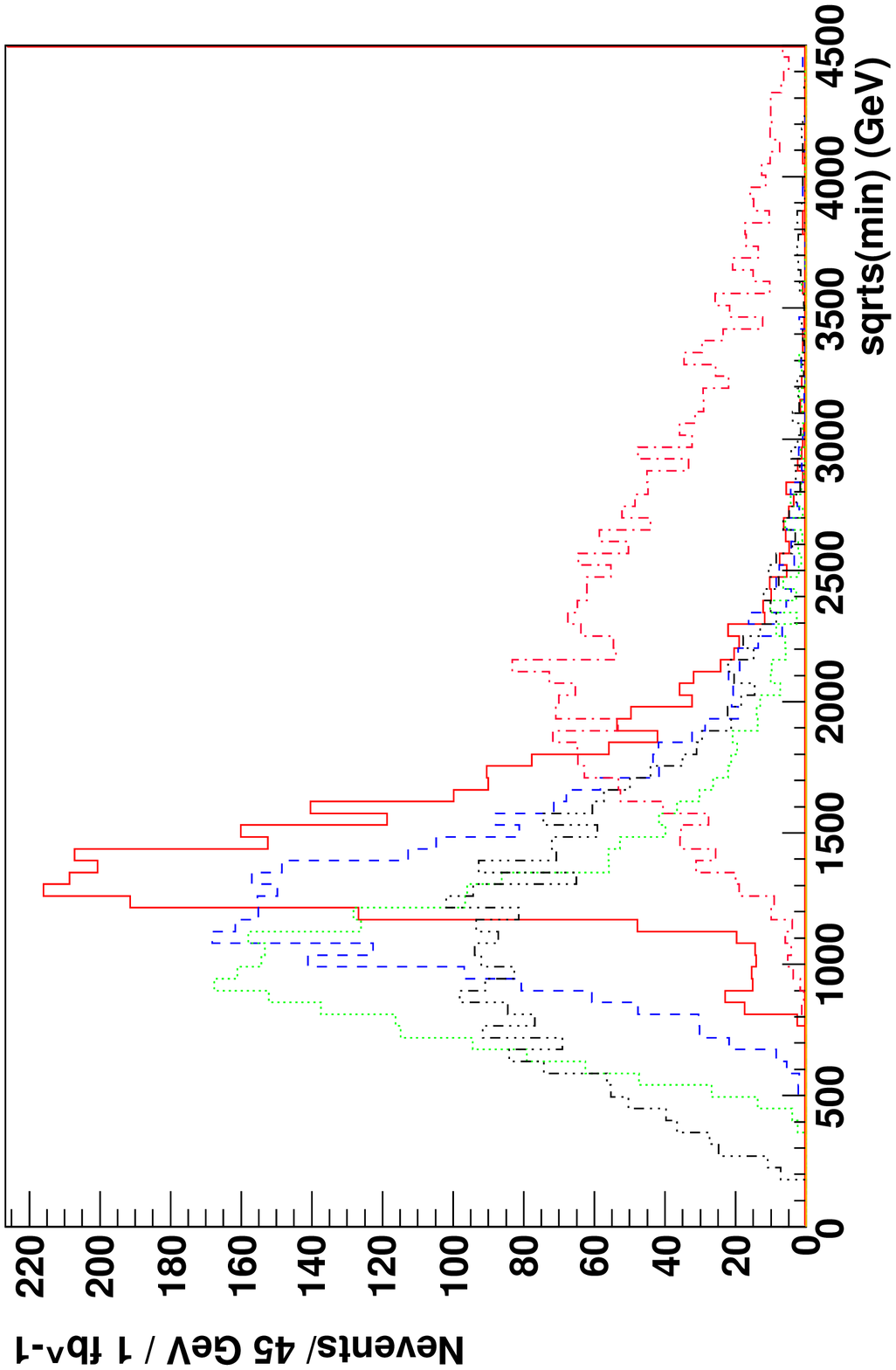}
\put(95,100){{ \footnotesize $\scriptstyle \sqrt{\hat{s}}_\text{part}$}}
\put(45,120){{ \footnotesize $\scriptstyle \sqrt{\hat{s}}_\text{min}^\text{part}$}}
\put(25,65){{\footnotesize $\scriptstyle \sqrt{\hat{s}}_\text{min}^\text{ana}$}}
\put(155,35){{ \footnotesize  $\scriptstyle \sqrt{\hat{s}}_\text{min}^{\text{cal, no } \eta}$ }}
\put(20,50){{\footnotesize $\scriptstyle \sqrt{\hat{s}}_\text{min}^\text{cal,$\eta$}$}}
%\put(0,100){huhu}
%\put(100,100){huhu2}
\end{overpic}
\end{center}
\end{minipage}
\caption{\label{fig:all3j} as Fig.~\ref{fig:allcol_noreq} $\int\mathcal{L}\,=\,1\,\fb^{-1}$: Left figure for exactly 0 leptons in the final state, 3 hardest jets, 14249 events), and shift between parton level and analysis level $\sqrt{\hat{s}}_\text{min}$ is about 100 \GeV. Right figure for exactly 2 leptons in the final state, 2 hardest jets (2745 events). Here, the shift between parton level and analysis level $\sqrt{\hat{s}}_\text{min}$ is about 200 \GeV~(reduces to 100 \GeV~if parton level tau vectors are used)}
\end{figure}

\begin{figure} 
\vspace{10mm}
\begin{minipage}{0.5\textwidth}
\begin{center}
\begin{overpic}[width=0.65\textwidth, angle=-90]{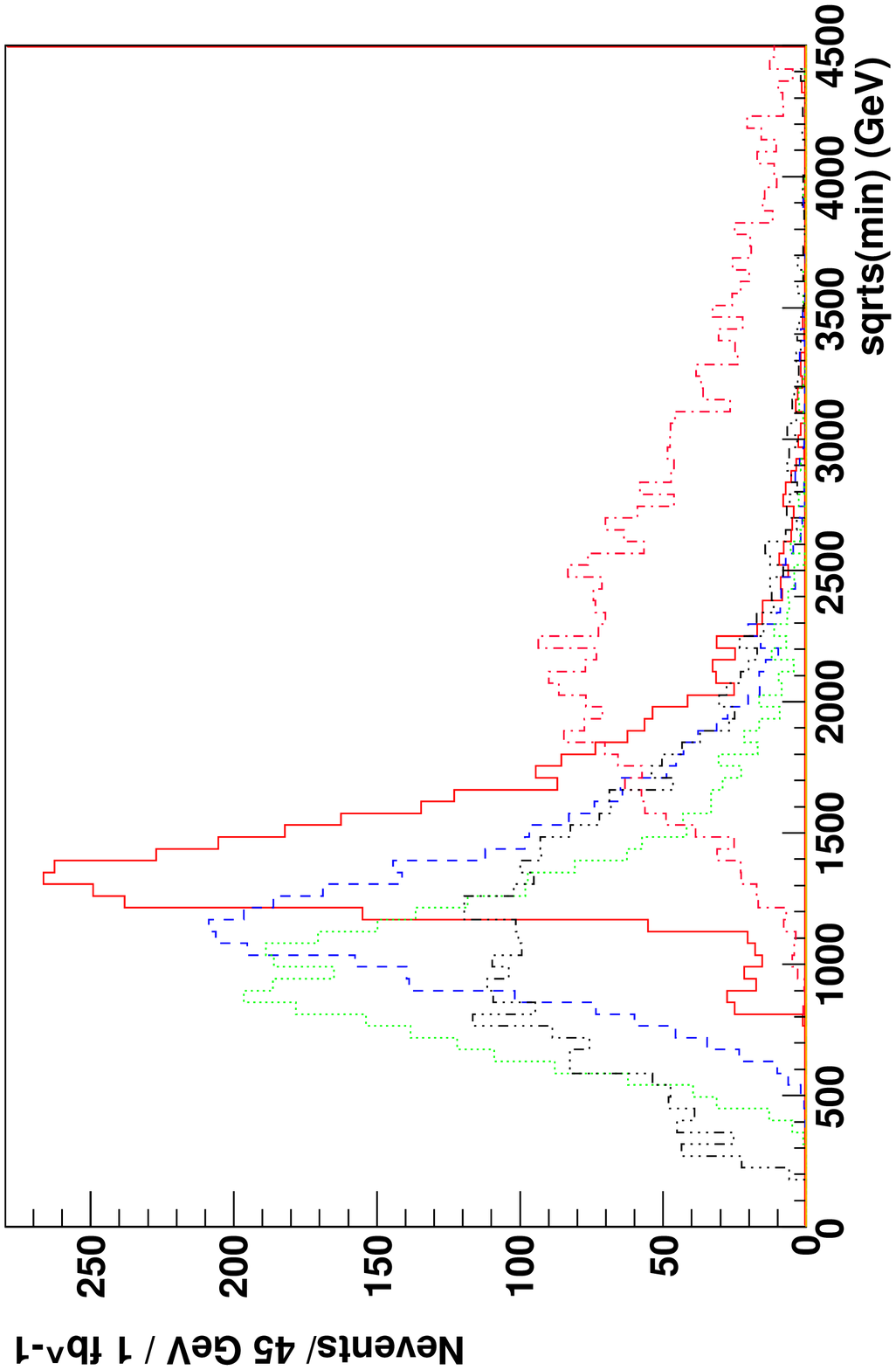}
%\put(0,100){huhu}
%\put(100,100){huhu2}
\put(90,100){{ \footnotesize $\scriptstyle \sqrt{\hat{s}}_\text{part}$}}
\put(45,115){{\footnotesize $\scriptstyle \sqrt{\hat{s}}_\text{min}^\text{part}$}}
\put(25,65){{\footnotesize $\scriptstyle \sqrt{\hat{s}}_\text{min}^\text{ana}$}}
\put(155,35){{\footnotesize  $\scriptstyle \sqrt{\hat{s}}_\text{min}^{\text{cal, no } \eta}$ }}
\put(20,40){{\footnotesize $\scriptstyle \sqrt{\hat{s}}_\text{min}^\text{cal,$\eta$}$}}
\end{overpic}
\end{center}
\end{minipage}
\begin{minipage}{0.5\textwidth}
\begin{center}
\begin{overpic}[width=0.65\textwidth, angle=-90]{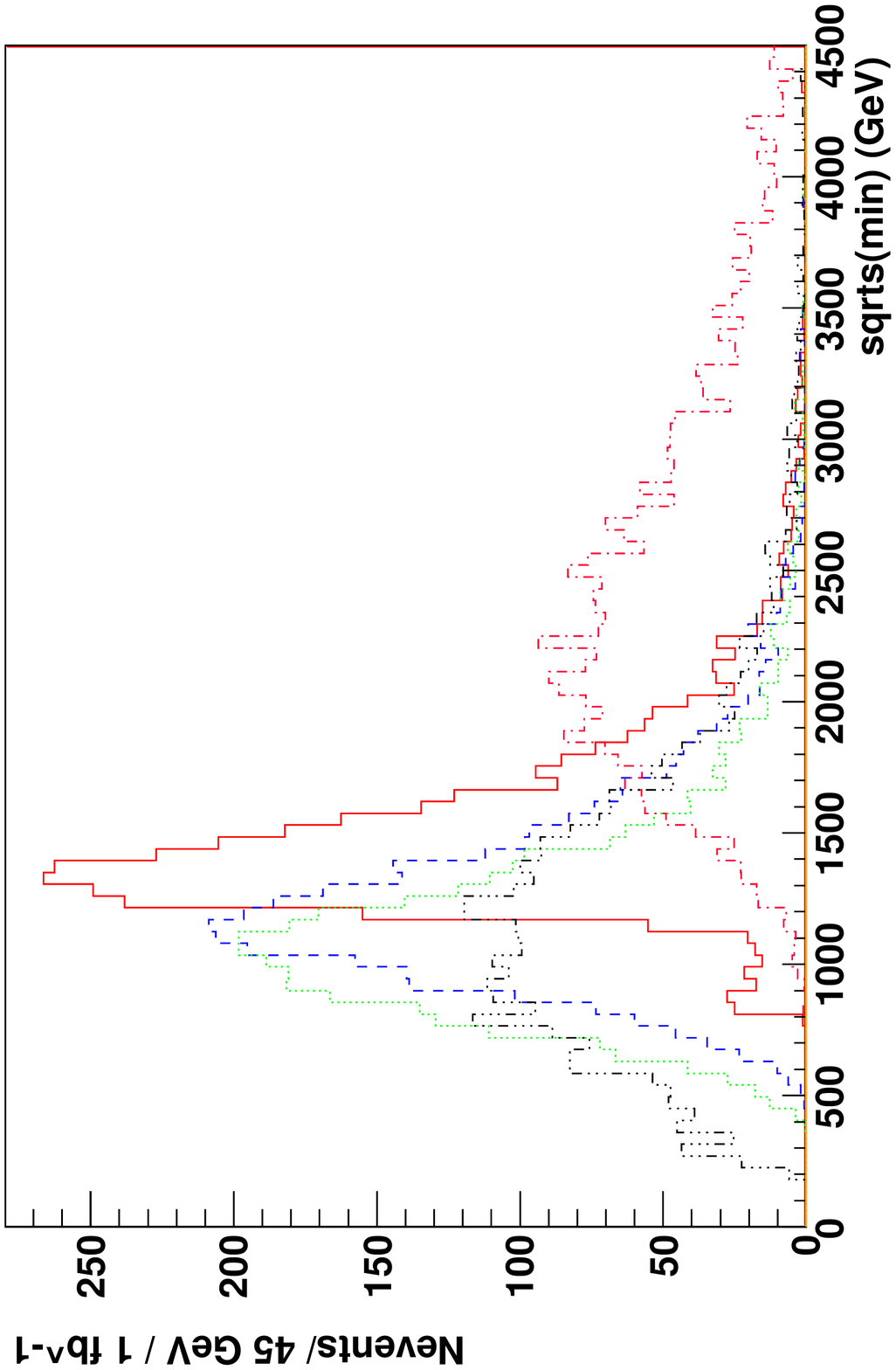}
\put(90,100){{ \footnotesize $\scriptstyle \sqrt{\hat{s}}_\text{part}$}}
\put(45,115){{ \footnotesize $\scriptstyle \sqrt{\hat{s}}_\text{min}^\text{part}$}}
\put(25,65){{ \footnotesize $\scriptstyle \sqrt{\hat{s}}_\text{min}^\text{ana}$}}
\put(155,35){{ \footnotesize  $\scriptstyle \sqrt{\hat{s}}_\text{min}^{\text{cal, no } \eta}$ }}
\put(20,40){{\footnotesize $\scriptstyle \sqrt{\hat{s}}_\text{min}^\text{cal,$\eta$}$}}
\end{overpic}
\end{center}
\end{minipage}
\caption{\label{fig:all_1l2j} as Fig.~\ref{fig:allcol_noreq} $\int\mathcal{L}\,=\,1\,\fb^{-1}$, exactly 1 tau lepton in the final state, 2 hardest jets (3260 events) Shift between parton level and analysis level peak is 170 \GeV (left) and reduces to 130 \GeV (right) when parton level tau vectors are used for analysis level objects.}
\end{figure}

\begin{figure} 
\begin{minipage}{0.5\textwidth}
\begin{center}
\begin{overpic}[width=0.60\textwidth, angle=-90]{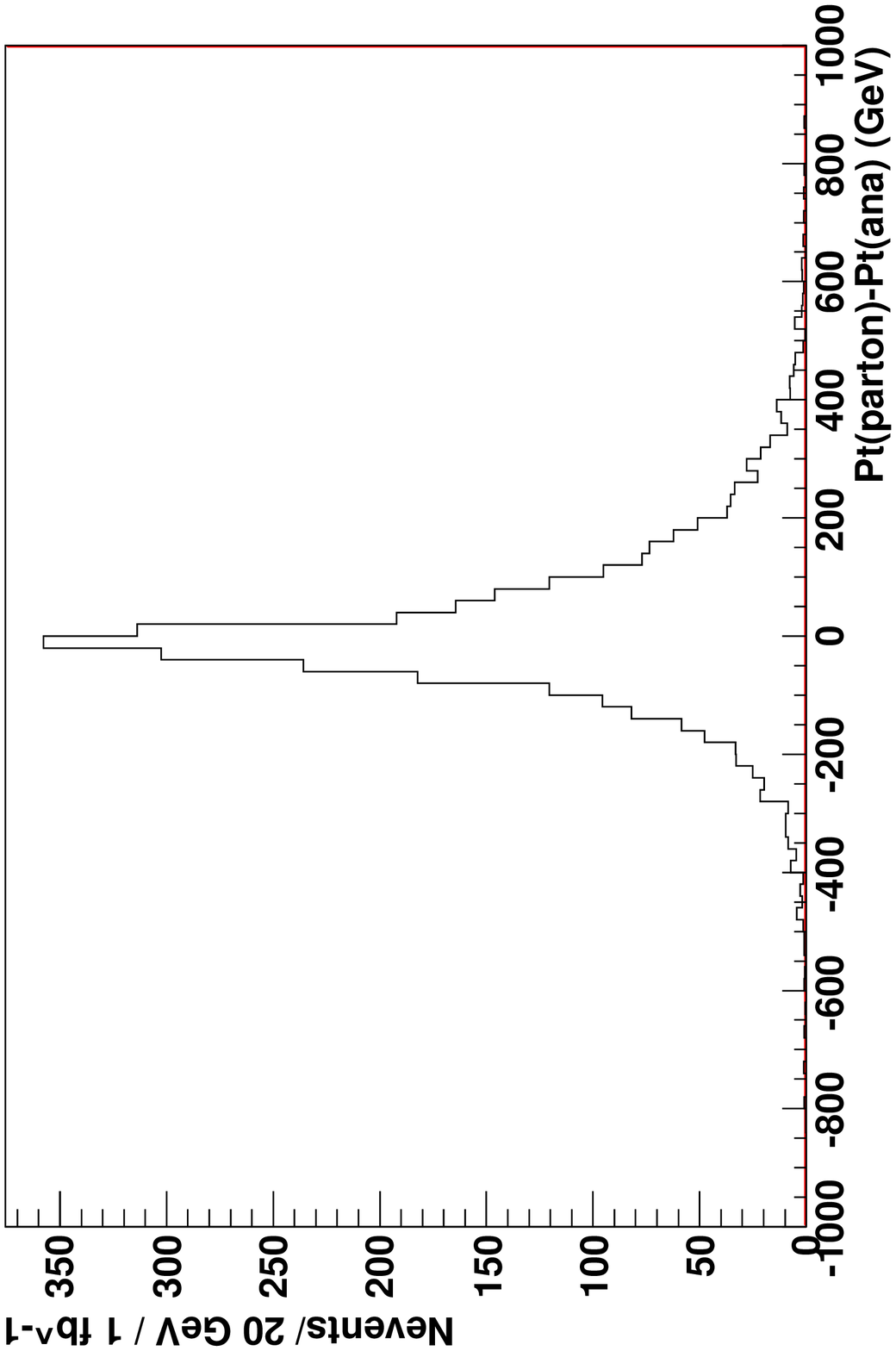}
\put(115,100){$\Delta P_T$}
\end{overpic}
\end{center}
\end{minipage}
\begin{minipage}{0.5\textwidth}
\begin{center}
\begin{overpic}[width=0.60\textwidth, angle=-90]{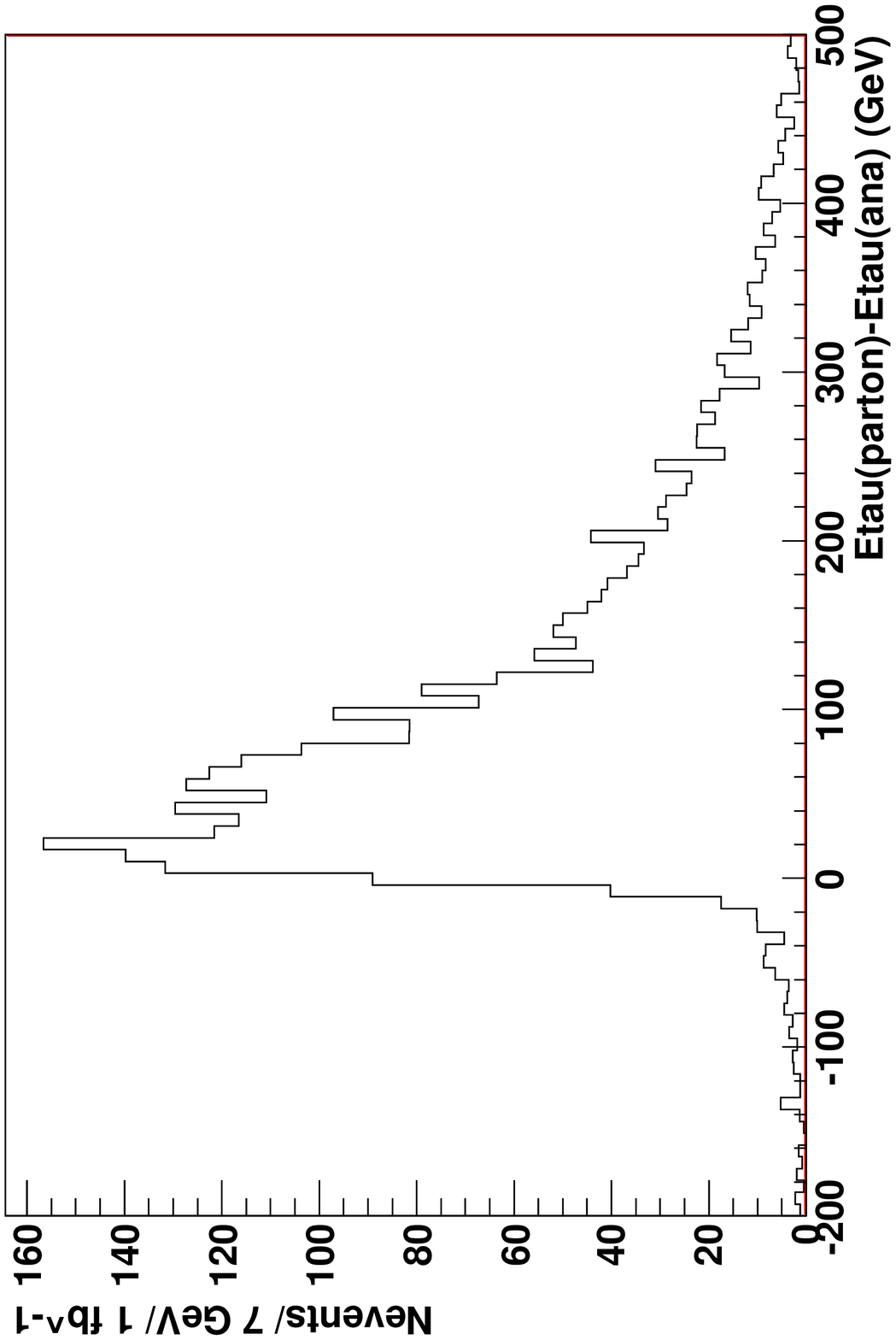}
%\put(130,50){$\Delta E_\tau$}
\put(95,100){$\Delta E_\tau$}
\end{overpic}
\end{center}
\end{minipage}
\caption{\label{fig:diffPt} Difference between parton level and analysis object level total transverse momentum (left) and tau jet energy (right) for the 1 tau 2 jet channel. While the shift between the two values for the transverse momentum peaks around around zero, the shift between the parton level tau and analysis level tau jet energy is quite large, due to the escaping neutrino in the tau jet reconstruction. The difference in the $P_Z$ distribution (not shown here) exhibits a similar behaviour as the $P_T$ distribution.}
\end{figure}

\begin{figure} 
\begin{minipage}{0.5\textwidth}
\begin{center}
\begin{overpic}[width=0.60\textwidth, angle=-90]{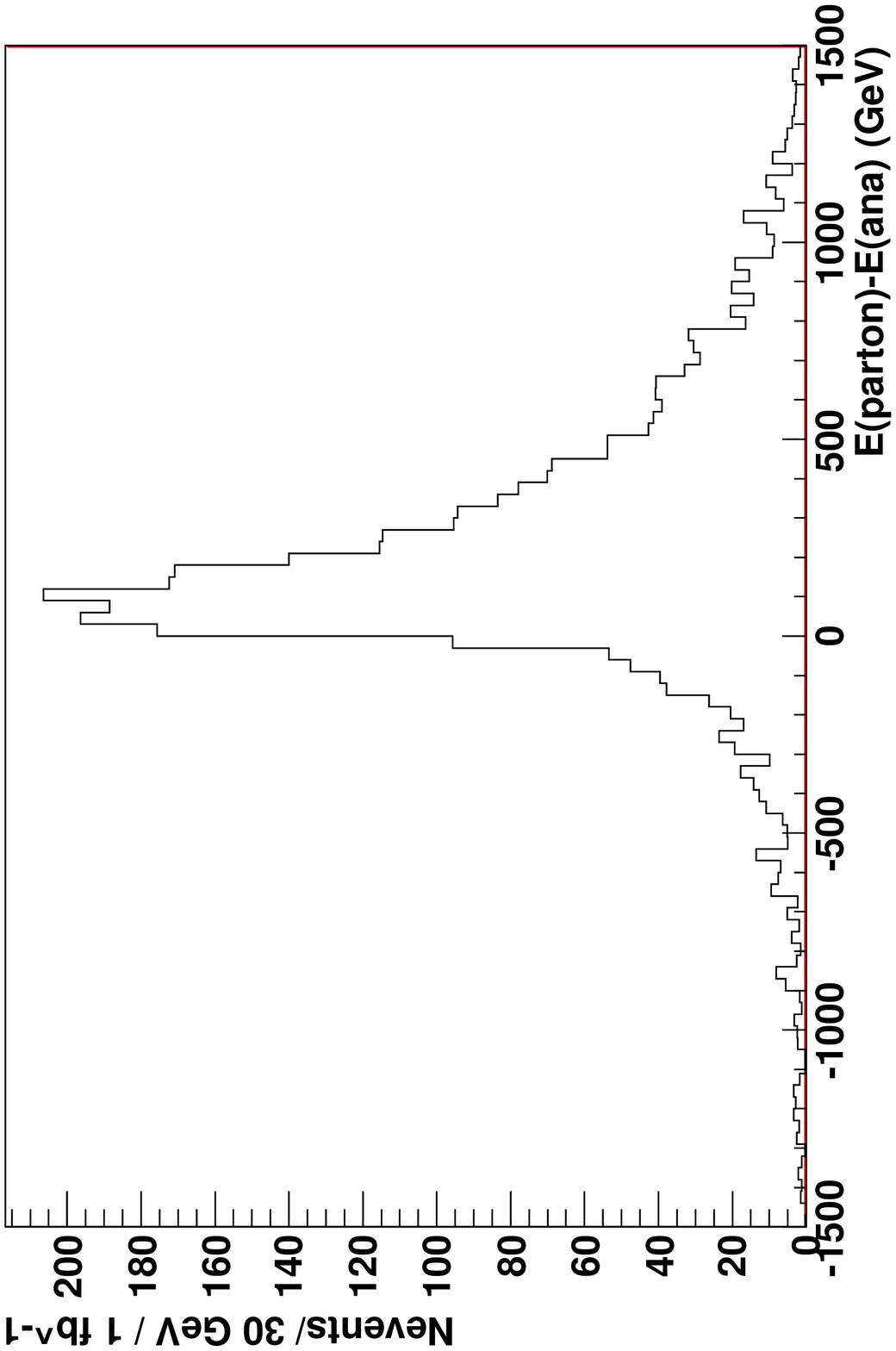}
\put(125,100){$\Delta E$}
\end{overpic}
\end{center}
\end{minipage}
\begin{minipage}{0.5\textwidth}
\begin{center}
\begin{overpic}[width=0.60\textwidth, angle=-90]{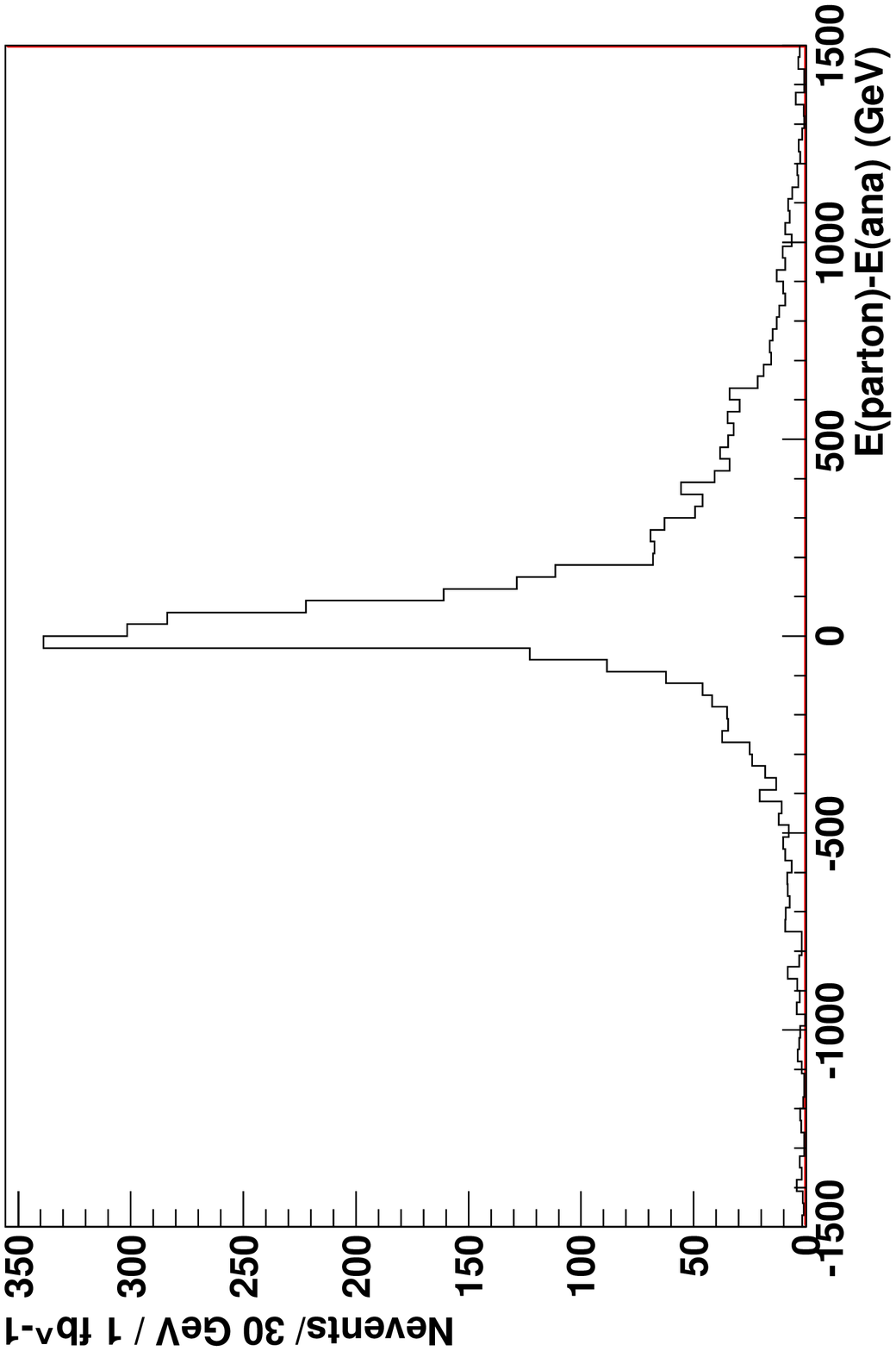}
\put(115,100){$\Delta E, {\scriptstyle \tau=\tau^\text{parton}}$}
\end{overpic}
\end{center}
\end{minipage}
\caption{\label{fig:diffE} Difference between parton level and analysis object level total visible energy for the 1 tau 2 jet channel. Left side shows the real difference, while on the right hand side analysis level tau jets were replaced by parton level taus. While we originally observe a large shift between the two values, originating from the escaping tau neutrinos on reconstruction object level and with a peak on the order of $\mO(200\,\GeV)$, the distribution of the difference peaks around zero when parton level tau four-vectors are used for the calculation of the analysis level observable.}
\end{figure}

Figures \ref{fig:all3j} and \ref{fig:all_1l2j} show the true $\sqrt{\hat{s}}$, parton level $\sqrt{\hat{s}}_\text{min}$, reconstruction level $\sqrt{\hat{s}}_\text{min}$ as well as the original calorimeter based variable with and without a cut in the magnitude of the pseudo rapidity $|\eta|\,<\,1.4$ for several explicit final states. We observe a similar behavior as in the overall sample, cf. Fig.~\ref{fig:allcol_noreq}: the parton level variable peaks around the actual production threshold, while there is a shift to lower peak values for the analysis level quantity. In order to understand the origin of this shift, we investigate this for a final state which initially exhibits a large difference between these quantities. We consider the 2 jet 1 tau-lepton channel, where originally the  $\sqrt{\hat{s}}_\text{min}$ peak positions differ by about $170\,\GeV$. From Eq.~(\ref{eq:sqrtshat_meas}), we see that the definition of $\sqrt{\hat{s}}_\text{min}$ depends on the following independently measured quantities:
\begin{\eqn*}
E_\text{vis},\,P_Z,\,|\ora{P}_T|¸\,=\,P_T\,=\,E_T\,=\,\slashed{E}_{T}.
\end{\eqn*}
In order to investigate the origin of the shift between the parton level and analysis level peak positions, we therefore consider each of these variables separately and plot the difference between the respective parton level and analysis level quantity; the results are shown in Figures \ref{fig:diffPt} and \ref{fig:diffE}. We observe 
 that, while the differences between parton and analysis level $P_{Z}, P_T$ basically peak around zero, there is an average discrepancy $\sim\,50\,-\,100\,\GeV$ between the parton and analysis level total visible energy. However, this discrepancy is accounted for by the fact that when changing from parton to hadron level, we replace the (visible) parton level tau by the (visible) tau-jet and the (invisible) tau-neutrino: 
\begin{\eqn*}
\tau^\text{part}\,\rightarrow\,\tau_\text{jet}+\nu_\tau.
\end{\eqn*}
In this transition, we equally shift the four momenta of the tau neutrinos from the visible to the invisible contribution of the definition of $\sqrt{\hat{s}}_\text{min}$ (cf.  Eq. (\ref{eq:sqrtshat_meas})): 
\begin{eqnarray*}
&&P_\text{vis}^\text{part}\,=\,...+p^\text{part}_\tau+...,\;M_\text{inv}^\text{part}\,\equiv\,M_\text{inv}\\
&&P^\text{ana}_\text{vis}\,=\,...+p_{\tau_\text{jet}}+...,\;M_\text{inv}^\text{ana}\,=\,M_\text{inv}^\text{part}+m_{\nu_{\tau}}.
\end{eqnarray*}
 We consider the neutrinos to be massless; therefore, we can leave the sum of all invisible particles' masses $M_\text{inv}$ unchanged. As the original variable definition of $\sqrt{\hat{s}}_\text{min}$ and the subsequent correlation in Eqn.~(\ref{eq:conj}) only depend on the heavy initial cascade particles, but not on the actual number of visible and invisible decay products, the observed change in the visible energy due to the escaping neutrinos at analysis level should then be compensated by associated changes in $P_Z,\,P_T$ on an event by event basis, leading to a similar peak behavior of the $\sqrt{\hat{s}}_\text{min}$ distribution at parton and analysis level.   
%which is reflected in the discrepancy between parton and analysis object level definition total taujet energy. 
% However, we here do not intend to provide an optimized tau reconstruction algorithm; therefore, 
Here, in order to assess the overall impact of this shift and a possible poor reconstruction of the tau decay products\footnote{Note that, in case the shift cannot be explained by poor reconstruction of the decay products alone, this equally opens the window to a possible {\sl topology}-dependence of $\sqrt{\hat{s}}_\text{min}$; we thank K.Sakurai for pointing this out.}, we perform a gedankenexperiment and change into a more ideal world where we idealistically reverse the analysis level tau-jet reconstruction and take the parton level tau four-vectors for the analysis level variable. In this case, the shift in the peak position of $\sqrt{\hat{s}}_\text{min}$ reduces to roughly $130\,\GeV$. An alternative though less sophisticated way to determine the peak position is to consider the bin which contains a maximal number of entries; using this definition of the distribution peak position, the original shift between parton level and analysis level $\sqrt{\hat{s}}_\text{min}$ reduces from $200\,\GeV$ to $90\,\GeV$ if the parton level tau vectors are used at analysis level. A similar study for the 2 tau lepton 2 jet channel, which originally equally exhibits a quite large shift between the peak positions, shows that the effect of tau misidentification is $\mO(100\GeV)$ for both peak position definitions, reducing to $\sim\,100\,\GeV$ in both cases when tau misidentification is removed.
%The 2 jet one lepton channel primarily exhibits a large shift of about $200\,\GeV$ between the parton level and reconstruction level variable; however, this shift reduces to about  $45\,\GeV$ if we take the  "correct" (= parton level) tau vectors for the analysis level variable, cf Figure \ref{fig:all_1l2j_truetaus}.
A similar effect can be observed for other specific final state signatures, cf. Table \ref{tab:finstats}. 
%\vspace{10mm}\\
%\begin{figure} 
%\begin{center}
%\includegraphics[angle=-90, width=0.6\textwidth]{plots/true_all_all_1tau3jets.eps}
%\end{center}
%\caption{as Fig. \ref{fig:all2j2l} $\int\mathcal{L}\,=\,1\,\fb^{-1}$, exactly 1 tau lepton in the final state, 3 hardest jets (not on calorimeter level though). Corresponds to 2535 events.}
%\end{figure}
\\
\begin{table}
\begin{\eqn*}
\begin{array}{c|c|c|c}
\text{final state}&\sqrt{\hat{s}}_\text{min}^\text{peak;parton}\,[\GeV]&\sqrt{\hat{s}}_\text{min}^\text{peak;ana}\,[\GeV]&\sqrt{\hat{s}}_\text{min}^\text{peak;ana,$\tau\,=\,\tau_{p}$}\,[\GeV]\\ \hline \hline
0\,l\,3\,j&1190\,\pm\,5&1072\,\pm\,6&1072\,\pm\,6\\
2\,l\,3\,j&1271\,\pm\,8&1128\,\pm\,8&1257\,\pm\,8\\
1\,\tau\,3\,j&1204\,\pm\,7&1123\,\pm\,8&1210\,\pm\,8\\
2\,l\,2\,j&1231\,\pm\,7&1001\,\pm\,7&1105\,\pm\,6\\
1\,\tau\,2\,j&1157\,\pm\,7&990\,\pm\,7&1031\,\pm\,8
\end{array}
\end{\eqn*}
\caption{\label{tab:finstats}Comparison of peak positions from Gaussian fits for parton level $\sqrt{\hat{s}}_\text{min}$ (${\scriptstyle \sqrt{\hat{s}}_\text{min}^\text{peak;parton}}$)  and analysis level $\sqrt{\hat{s}}_\text{min}$ (${\scriptstyle \sqrt{\hat{s}}_\text{min}^\text{peak;ana}}$) for specific final states, specified by the number of visible final state leptons ($l$), jets ($j$), and $\tau$-leptons ($\tau$), corresponding to dominant decay chains in the complete SPS1a sample. Values for the peak position of the analysis level quantity with perfect tau jet reconstruction (${\scriptstyle \sqrt{\hat{s}}_\text{min}^\text{peak;ana,$\tau\,=\,\tau_{p}$}}$) are also given. For most final states, the effect of the peak shift due to imperfect tau jet reconstruction is $\mO(100\,\GeV)$.}
\end{table}

Although we still obtain a quite large shift for specific final state
signatures, we have seen that, when using parton level taus for the analysis level observable and therefore suppressing possible effects from poor tau reconstruction, the inclusive sample peaks at the same
value for both parton and analysis level  $\sqrt{\hat{s}}_\text{min}$
distributions, cf. Fig.~\ref{fig:true_all_truetaus}. We therefore
conclude that, with correctly identified analysis level objects, the
peak position of the parton level $\sqrt{\hat{s}}_\text{min}$ can
indeed be reconstructed from generator level measurements\footnote{We
want to point out that the reconstruction level objects in Table
\ref{tab:objects}, through their definition by $p_T$ and $\eta$ cuts,
still depend on these two parameters; therefore,
a recovery of the hard scale from reconstruction level objects will always be obstructed by an implicit
dependence on the cut values in the analysis object definitions. However, in contract to the
calorimeter-based variable and the cut in pseudorapidity originally
proposed in \cite{Konar:2008ei}, we here use object level definitions
which are more optimized to the reconstruct the hard scattering
event. 
%Indeed {\sl all} physics analyses which are done on analysis
%level exhibit the
%implicit dependecies on cut values in the reconstruction object
%definitions. 
We thank B. Webber for bringing this point to our attention.}; however, we want to emphasize that the {\sl correlation} between the peak position and the actual heavy particle production threshold only exists in the form of a conjecture which lacks a rigorous proof. In case the conjecture proves to hold in all cases, the analysis level $\sqrt{\hat{s}}_\text{min}$ variable indeed gives a quite easy grasp on the threshold of the new physics pair-produced particles. Although this analysis was done in a specific scenario, where only certain initial heavy particle spin states are allowed, we saw that our conclusions hold for all possible spin combinations we considered. As our study relies on purely kinematic variables, we are therefore confident that these also hold for other spin combinations both for the heavy initial pair-produced particles as well as the particles in the decay chains, i.e. especially for other (also non-SUSY) BSM scenarios. 
\subsection{Comment on additional soft physics effects}
The data set used in this study contains soft physics in the form of initial and final state radiation as described in Section \ref{sec:data}, but no simulation of underlying event or pileup. However, the criticism which was expressed by the authors of \cite{Papaefstathiou:2009hp} exactly concerns the dependence of soft physics in terms of ISR, which has been addressed in this work. Additionally, soft physics can enter in the form of minimum bias events, underlying event and pileup. We believe that the cuts in Table \ref{tab:objects} are sufficiently hard enough to suppress minimum bias events (\cite{privcomm}, as well as section 6.1 in \cite{Moraes:2007rq})). Underlying event as well as pileup effects can still distort the overall result for the peak position; however, we believe that these issues should be pursued in an experimental study, in combination with a collaboration internal full detector simulation. We can give a first estimate of the effect of underlying event fake $P_T$ contributions by adding $\Delta P_T^\text{fake}\,=\,10\,\GeV$, which corresponds to a conservative upper limit of the average $P_T$ from the underlying event \cite{Moraes:2007rq, Tricoli:2009zz}, in the definition of $\slashed{E}_T$ in Eq. (\ref{eq:sqrtshat_meas}); in this case, the best fit value from $400\,\GeV\,\leq\,\sqrt{\hat{s}}_\text{min}\,\leq\,1400\,\GeV$ is again given by 
\begin{eqnarray*}
%\text{analysis level, fake $\Delta \slashed{P}_T^\text{fake}$ from UE:  } 
\sqrt{\hat{s}}_\text{min}^\text{peak}(\Delta P^\text{fake}_T\,=\,10\,\GeV)&:&(1082\,\pm\,4)\,\GeV
\end{eqnarray*}
which completely agrees with the value without the addition of $\Delta P_T^\text{fake}$. Changing the fake additional transverse momentum to $\Delta P^\text{fake}_T\,=\,100\,\GeV$ leads to the result\footnote{Simulations using more recent tunes for a $14\,\TeV$ LHC, as eg the Pythia C4 tune \cite{Corke:2010yf}, point to additional $\Delta P^\text{fake}_T\,\sim\,30-40\,\GeV$ for similar/ less stringent object definition values \cite{deepak}; however, following the above considerations the induced error then is certainly $\sim\,\GeV$, which again corresponds to a relatively small uncertainty in the determination of $\sqrt{\hat{s}}_\text{min}^\text{peak}$ .}
\begin{eqnarray*}
%\text{analysis level, fake $\Delta \slashed{P}_T^\text{fake}$ from UE:  } 
\sqrt{\hat{s}}_\text{min}^\text{peak}(\Delta P^\text{fake}_T\,=\,100\,\GeV)&:&(1093\,\pm\,4)\,\GeV.
%\sqrt{\hat{s}}_\text{min}^\text{peak}(\Delta P_t\,=\,200\,\GeV)&:&(1128\,\pm\,4)\GeV
\end{eqnarray*}
In fact, as $M_\text{inv}$ and $\slashed{E}_T$ appear in the same form in the definition of $\sqrt{\hat{s}}_\text{min}$, the generic effects of additional fake $P_T$s can be estimated from Figure \ref{fig:allparton}. 
From underlying events,  $\Delta P^\text{fake}_T\,\lesssim \,100\,\GeV$, and we therefore estimate the uncertainty related to underlying event fake transverse momentum to be generically much smaller than the tau reconstruction effects discussed above. A more realistic investigation of these experimentally dominated effects, which should include a full detector simulation, is beyond the scope of this work\footnote{In fact the experimental collaborations are already applying algorithms to subtract $E_T$ due to underlying event; cf eg \cite{ATLAS-CONF-2010-056}. We thank S. Wahrmund for bringing this to our attention.}.
\section{SM background}\label{sec:smbkg}
\begin{figure} 
\begin{minipage}{0.5\textwidth}
\begin{center}
\begin{overpic}[width=0.60\textwidth, angle=-90]{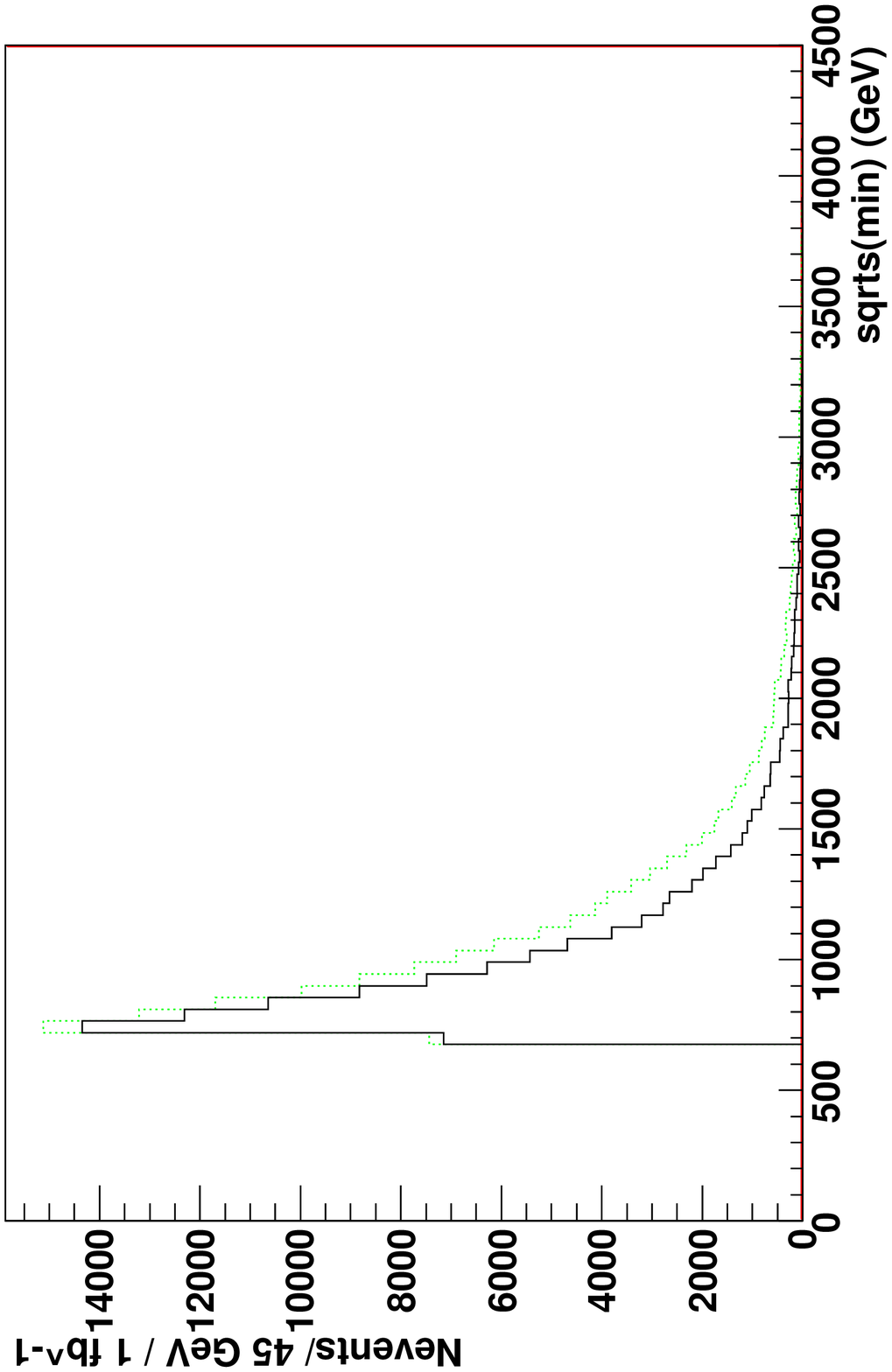}
%\put(130,50){$\Delta E_\tau$}
\put(70,100){$\scriptstyle { \sqrt{\hat{s}}_\text{min}^\text{(SM+BSM)}}\,{\text{(green;dotted)}}$}
\put(70,80){$\scriptstyle { \sqrt{\hat{s}}_\text{min}^\text{SM}}\,{\text{(black; solid)}}$}
\end{overpic}
\end{center}
\end{minipage}
\begin{minipage}{0.5\textwidth}
\begin{overpic}[width=0.60\textwidth, angle=-90]{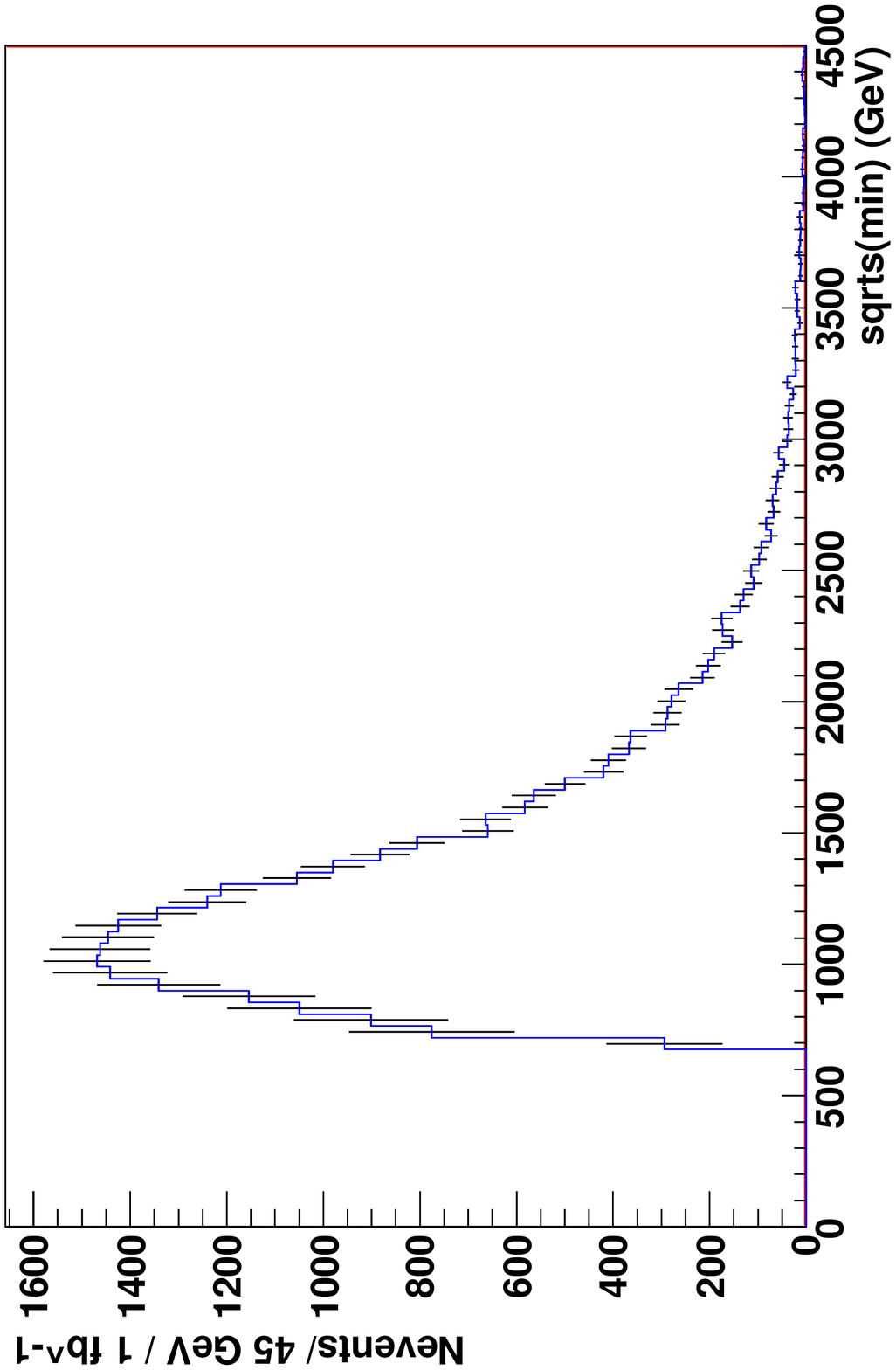}
%\put(130,50){$\Delta E_\tau$}
\put(90,100){$\scriptstyle  \sqrt{\hat{s}}_\text{min}^\text{(SM+BSM)-SM}$}
\end{overpic}
\end{minipage}
\caption{\label{fig:smvsall_4bkg} Analysis level $\sqrt{\hat{s}}_\text{min}$ after a cut $\sqrt{\hat{s}}_\text{min}> 700\,\GeV$. Dominant six SM backgrounds after cut ($W+2\,j,\,W+3\,j,W+4\,j,\,t\bar{t},\,t\bar{t}+1\,j,\,t\bar{t}+2\,j$) are included. Left: SM+BSM (green, dotted; 136834 events) and SM only (black, solid; 108017 events). In the sum and without further suppression cuts, the peak structure disappears. Right: Difference between (SM+BSM) and (SM). Assuming the SM background is well-known, the peak structure of the BSM signal is recovered. The difference is much larger than the statistical error. \vspace{3mm}}
\end{figure}

\begin{figure} 
\begin{minipage}{0.5\textwidth}
\begin{center}
\begin{overpic}[width=0.60\textwidth, angle=-90]{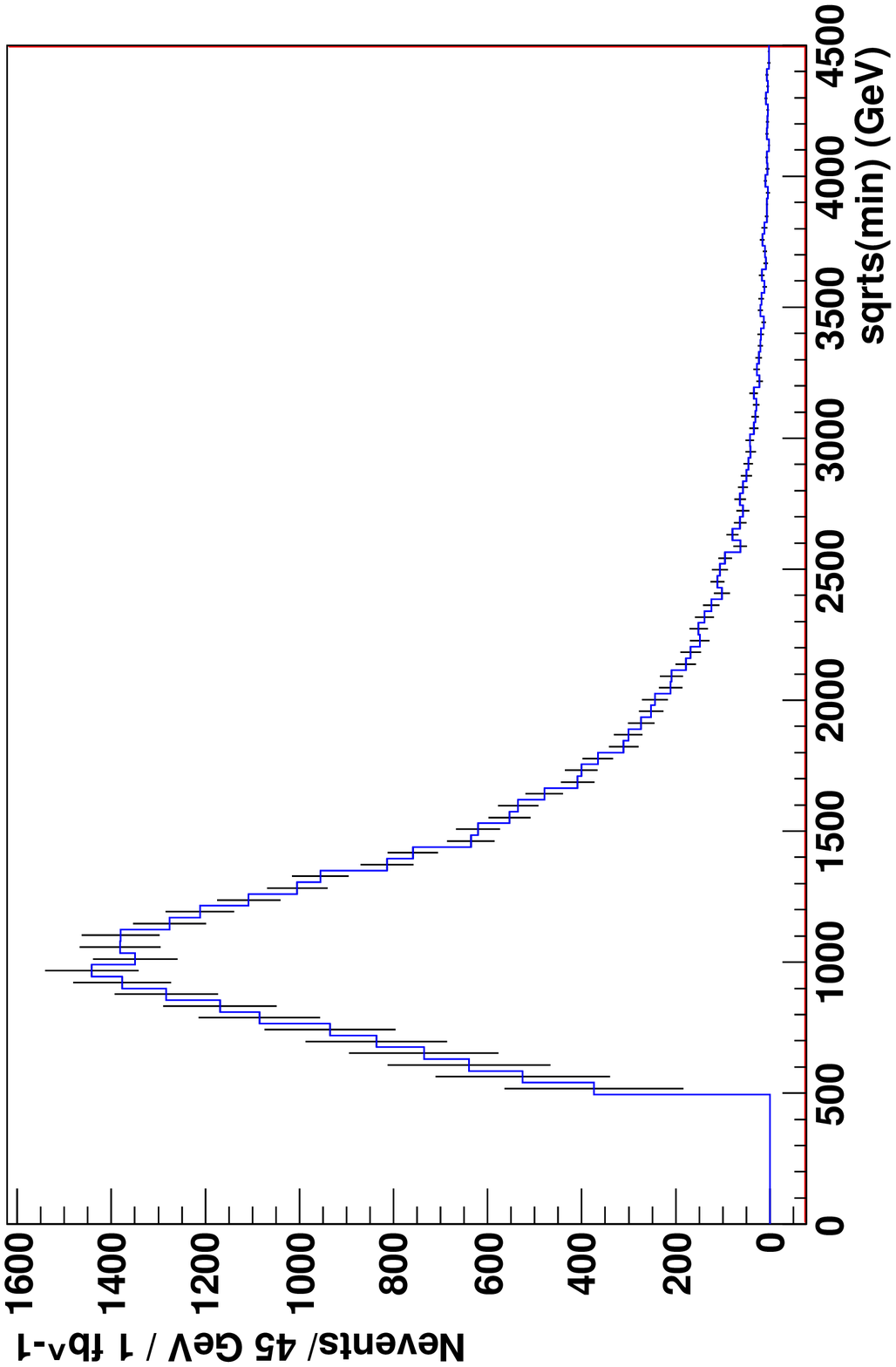}
%\put(130,50){$\Delta E_\tau$}
\put(90,100){$\scriptstyle  \sqrt{\hat{s}}_\text{min}^\text{(SM+BSM)-SM} $}
\put(90,80){$\scriptstyle M_\text{inv}=\,0\,\GeV$}
\end{overpic}
\end{center}
\end{minipage}
\begin{minipage}{0.5\textwidth}
\begin{center}
\begin{overpic}[width=0.60\textwidth, angle=-90]{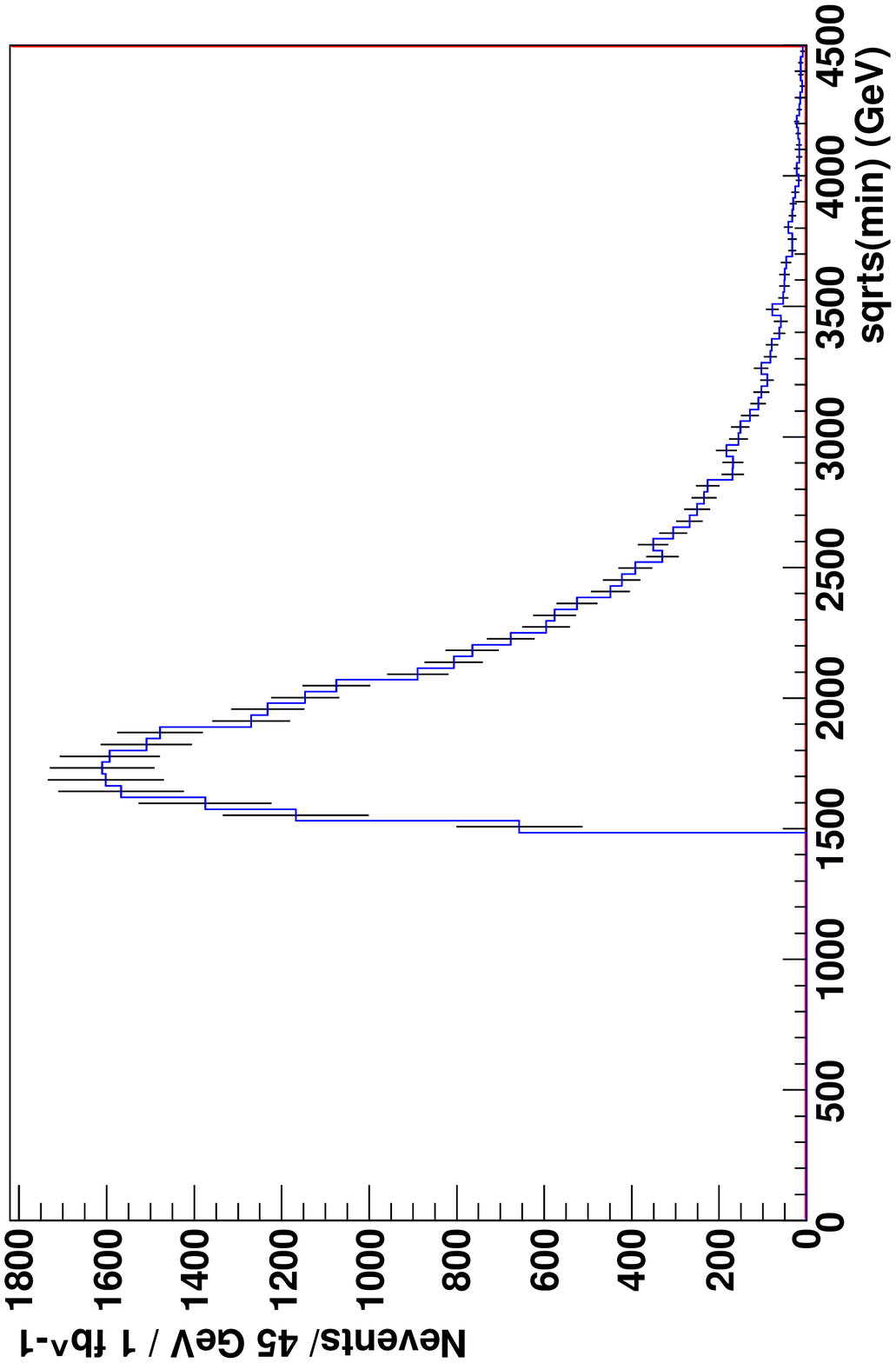}
%\put(130,50){$\Delta E_\tau$}
\put(120,100){$\scriptstyle  \sqrt{\hat{s}}_\text{min}^\text{(SM+BSM)-SM}$}
\put(120,80){$\scriptstyle M_\text{inv}=1000\,\GeV$}
%\put(105,100){$\Delta E_\tau$}
\end{overpic}
\end{center}
\end{minipage}
\caption{\label{fig:diff_4bkg_0} Difference between (BSM+SM) and (SM) for $M_\text{inv}\,=\,0\,\GeV$ and a cut $\sqrt{\hat{s}}_\text{min} > 500\,\GeV$ (left; 29815 events) as well as $M_\text{inv}\,=\,1000\,\GeV$ and a cut $\sqrt{\hat{s}}_\text{min} > 1500\,\GeV$ (right; 27802 events).   Assuming the SM background is well-known, the peak structure of the BSM signal is recovered. The difference is much larger than the statistical error. }
\end{figure}

In this section, we investigate $\sqrt{\hat{s}}_\text{min}$ when SM background is included, as well as its use for the reduction of SM background in new physics searches. As an example, we consider $W\,+$ jets and $t\bar{t}\,+$ jets background. Due to the large cross sections, we applied an additional $\slashed{P}_T$ filter in the generation of the SM data sample, cf. Table \ref{tab:addfilter}. The cross sections after these additional cuts are given in Table \ref{tab:smxsecs}.
\begin{table}
\begin{\eqn*}
\begin{array}{c|c}
n_\text{leptons}&|\slashed{P}_{T,\text{min}}|\\ \hline
< 2&80\,\GeV\\
2&40\,\GeV\\
> 2&0\,\GeV
\end{array}
\end{\eqn*}
\caption{\label{tab:addfilter} Additional filters on magnitude of the total missing transverse momenum $|\slashed{P}_T|$ applied for SM background generation, depending on the number of final state leptons $n_\text{leptons}$. Leptons are required to obey the cut criterium $|p_T|\,>\,5\,\GeV$ for the magnitude of the transverse momentum and $|\eta|\,<\,3.2$ for the magnitude of pseudorapidity.}
\end{table}
\begin{table}
\begin{\eqn*}
\begin{array}{c|c|c}
n_\text{jets}&W+n\, \text{jets}&t\bar{t}+n\,\text{jets}\\ \hline
0&&75.2\\
1&&48.5\\
2&188&20.0\\
3&53.6&6.3\\
4&12.6&1.4
\end{array}
\end{\eqn*}
\caption{\label{tab:smxsecs} Cross sections in \pb~ for SM background with Alpgen; filters in Tab. \ref{tab:addfilter} were applied in the generation stage.}
\end{table}
As before, we investigate the peak position for
$\sqrt{\hat{s}}_\text{min}$ at parton and analysis level when using
the true BSM value  $M_\text{inv}\,=\,2\,m_{\wt{\chi}^0_1}$.  From Fig.
\ref{fig:smvsall_4bkg} we see that in the total number of events after
the cuts the peak structure disappears when no further SM cuts are
applied. However, assuming an accurate enough (data or Monte Carlo
driven) background subtraction, the peak structure is clearly visible
again and much larger than the statistical error,
%\footnote{We want to
%  note that systematic errors, which we have not included in our study, can of course
%  be larger than the ``pure'' BSM signal which has been obtained from
%  SM subtraction, especially for higher values of
%  $\sqrt{\hat{s}}_\text{min}$; we thank K. Sakurai for comments on
%  this point.}
 cf. Fig.~\ref{fig:smvsall_4bkg}. 
%We therefore conclude that, at least for the parameter point studied here, with SM background being well-known,  $\sqrt{\hat{s}}_\text{min}$ can be used as a BSM discovery variable and that for a true input value of $M_\text{inv}$, the scale of the new physics can be derived from the peak of both parton level and (properly defined) analysis level quanitites.\\
%Finally, we investigate whether the same effect can be seen if we vary $M_\text{inv}$, ie we do not necessarily use its {\sl true} value $M_\text{inv}\,=\,2\,m_{\wt{\chi}^{0}_{1}}$:
A similar behavior is observed when we vary the input variable
$M_\text{inv}$: Fig.~\ref{fig:diff_4bkg_0} shows the behavior for
$M_\text{inv}\,=\,0\,\GeV,\,\,1000\,\GeV$ respectively after SM
background subtraction; we see we obtain a clear BSM signal. We therefore conclude that, at least for the parameter point studied here, with SM background being well-known,  $\sqrt{\hat{s}}_\text{min}$ can be used as a BSM discovery variable and that for a true input value of $M_\text{inv}$, the scale of the new physics can be derived from the peak of both parton level and (properly defined) analysis level quantities.\\
\\

We furthermore assess the use of $\sqrt{\hat{s}}_\text{min}$ as a cut variable for SM background suppression. For this, we investigate the position of the peak for the different SM background channels considered here, where we again use $M_\text{inv}\,=\,2\,m_{\wt{\chi}^0_1}$.  The respective values are given in Table \ref{tab:smpeaks}. For a first estimate of these positions, we do not need to perform a more sophisticated fit, and we therefore follow the simplified approach by defining the peak positions according to the bin which has the maximal number of entries. 
\begin{table}
\begin{\eqn*}
{\textstyle
\begin{array}{c|c|c|c|c|c}
\text{process}&\sigma&\sqrt{\hat{s}}_\text{min}^\text{maxbin;parton}&\sqrt{\hat{s}}_\text{min}^\text{maxbin;analysis}&\sigma, {\scriptstyle \sqrt{\hat{s}}_\text{min}>700\GeV}&\sigma, {\scriptstyle \sqrt{\hat{s}}_\text{min}>800\GeV}\\
&[\pb]&[\GeV]&[\GeV]&[\pb]&[\pb]\\ \hline\hline
W+2\,j&188&450&405&29.03&19.07\\
W+3\,j&53.6&630&585&23.47&17.51\\
W+4\,j&12.6&900&765&9.03&7.63\\ \hline
\sum\,W+\text{jets}&254.2&&&61.53&44.21\\ \hline
&&&&&\\
t\bar{t}&75.2&540&450&11.50&3.35\\
t\bar{t}+1\,j&48.5&675&585&20.71&14.41\\
t\bar{t}+2\,j&20.0&900&720&14.28&11.68\\
t\bar{t}+3\,j&6.3&1215&900&5.31&4.85\\
t\bar{t}+4\,j&1.4&1530&1215&1.29&1.24\\ \hline
%\text{SM}&405.6\,(659.8)&&&115.52\,(178.05)&79.74\,(123.95)\\ \hline
\sum\,t+\text{jets}&151.4&&&53.09&35.53\\ \hline
&&&&&\\
\tilde{q}\tilde{q}&6.56&1080&1035&5.59&5.03\\
\tilde{g}\tilde{g}&4.53&1260&1170&4.38&4.20\\
\tilde{g}\tilde{q}&19.96&1170&1035&18.85&17.82\\ \hline
\text{BSM}&31.05&&&28.82&27.05\\ \hline
%&&&&&\\
%S/B&0.077\,(0.047)&&&0.25\,(0.16)&0.33\,(0.22)
\end{array}
}
\end{\eqn*}
\caption{\label{tab:smpeaks} Cross sections $\sigma$ for SM background processes and $\sqrt{\hat{s}}_\text{min}$ maximal bin positions for parton level ($\scriptstyle \sqrt{\hat{s}}_\text{min}^\text{maxbin;parton}$) and analysis level ($\scriptstyle \sqrt{\hat{s}}_\text{min}^\text{maxbin;analysis}$) with standard analysis object definitions only; $M_\text{inv}\,=\,2\,m_{\wt{\chi}^0_1}$. Last two columns give cross sections $\sigma$ after $\sqrt{\hat{s}}_\text{min}$ cuts respectively. After a minimal analysis level cut on $\sqrt{\hat{s}}_\text{min}$, the $W$ and $t\bar{t}$ backgrounds are reduced by factors $3-6$, while we maintain roughly $90\,\%$ of the BSM signal.}
\end{table}
We see that the most dominant SM background channels have distribution peaks around $500-700\,\GeV$, while the BSM signals peak at higher values. We therefore apply two different cuts of $\sqrt{\hat{s}}_\text{min}\,\geq\,700\,\GeV$ and $\sqrt{\hat{s}}_\text{min}\,\geq\,800\,\GeV$ on all samples; the cross sections after these cuts are summarized in Table \ref{tab:cuteff}. We see that, for both cut values, while we only cut out around $10\%$ of the BSM signal, the dominant SM channels are suppressed by factor 3-6. We therefore conclude that $\sqrt{\hat{s}}_\text{min}$ can easily be used as a variable for SM background suppression, even for {\sl wrong} guesses for the total invisible mass $M_\text{inv}$. In the previous sections, we discussed how in our sample the peak of the $\sqrt{\hat{s}}_\text{min}$ variable is correlated with the real threshold for the hard production cross sections {\sl only if} the correct value input for $M_\text{inv}$ is used, cf. Eqn. (\ref{eq:conj}). For the background suppression, however, a correct guess or estimate of this value from other sources is not necessary, and we equally obtain a good SM background suppression with wrong input values for $M_\text{inv}$\footnote{We want to remind the reader that the {\sl same} value of $M_\text{inv}$ for the calculation of $\sqrt{\hat{s}}_\text{min}$ needs to be used in both SM and BSM samples; the correlation with the threshold however only holds for the sample with the equivalent {\sl correct} $M_\text{inv}$. We thank K. Matchev for reemphasizing this point.}.\\  

%As we did not study all SM backgrounds, we also list the separate suppression factors for different $M_\text{inv}$ assumptions in table \ref{tab:cuteff}.
\begin{table}\small
\begin{\eqn*}
{\scriptstyle
\begin{array}{c||c||c|c|c|c||c|c||c|c}
&&& & & &{ M_\text{vis, min}}& &{M_\text{eff, min}} &\\
&\text{no cut}&{ M_\text{inv}}=& & & &[\GeV]& &[\GeV] &\\
&&{2\,m_{\wt{\chi}^{0}_{1}}}&0\,\GeV&400\,\GeV&10^3\,\GeV&400& 500&400& 500
\\ \hline \hline
{\scriptstyle W+\text{ jets}}&254.2&62.53&81.13&48.73&46.73&70.29&42.41&75.89&43.3\\
{\scriptstyle t\bar{t}+\text{ jets}}&151.4&52.99&64.67&51.77&50.14&64.63&46.3&63.98&40.12\\
\text{BSM}&31.05&28.82&29.82&28.42&27.80&27.07&24.43&29.99&28.72
\end{array}
}
\end{\eqn*}
\caption{\label{tab:cuteff} Total cross sections $\sigma\,[\pb]$ for BSM as well as $W+\text{ jets}$  and $t\bar{t}+\text{ jets}$ backgrounds, without and with several cuts on different inclusive quantities. First column: no cut; second to fifth column: values for $\sqrt{\hat{s}}_\text{cut}\,=\,M_\text{inv}+500\,\GeV$, with varying $M_\text{inv}$ values; last four columns: cuts on $M_\text{vis}$ {\small (Eq. (\ref{eq:M}))} and $M_\text{eff}$ {\small (Eq. (\ref{eq:meff}))} respectively, where $M_\text{inv}\,=\,2\,m_{\wt{\chi}^0_1}$. While the maximal suppression factor for the BSM signal is around 1.11, the SM backgrounds are suppressed by factors $2-5$. Equal results, however, can easily be obtained by a cut on $M_\text{vis}$ or $M_\text{eff}$ . }
\end{table}
\subsection{Comparison with other (transverse) variables}
\begin{figure} 
\begin{minipage}{0.5\textwidth}
\begin{center}
\begin{overpic}[width=0.60\textwidth, angle=-90]{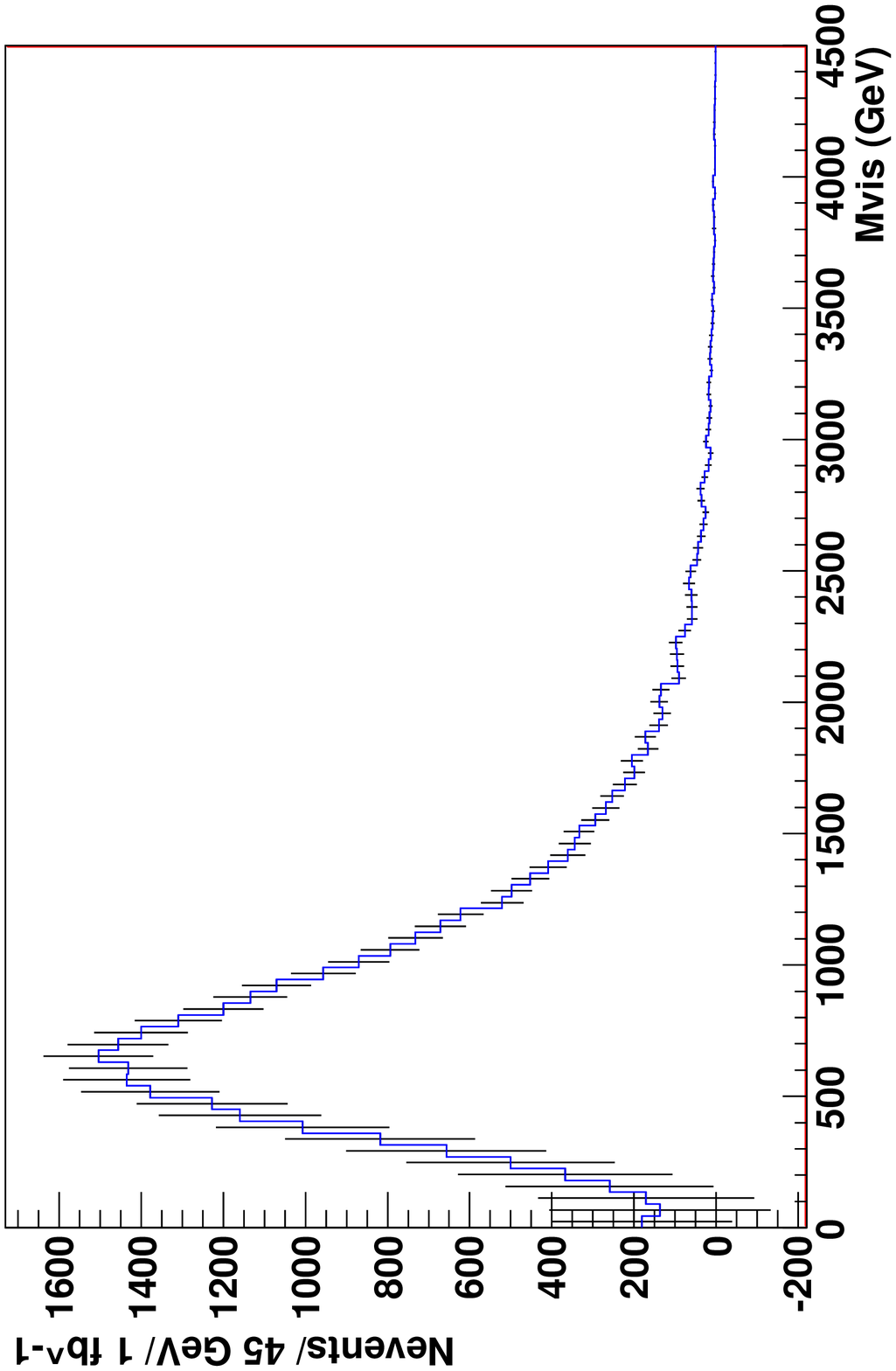}
%\put(130,50){$\Delta E_\tau$}
\put(80,100){$\scriptstyle M_\text{vis}^\text{(SM+BSM)-SM} $}
%\put(90,80){$\scriptstyle M_\text{inv}=0$}
\end{overpic}
\end{center}
\end{minipage}
\begin{minipage}{0.5\textwidth}
\begin{center}
\begin{overpic}[width=0.60\textwidth, angle=-90]{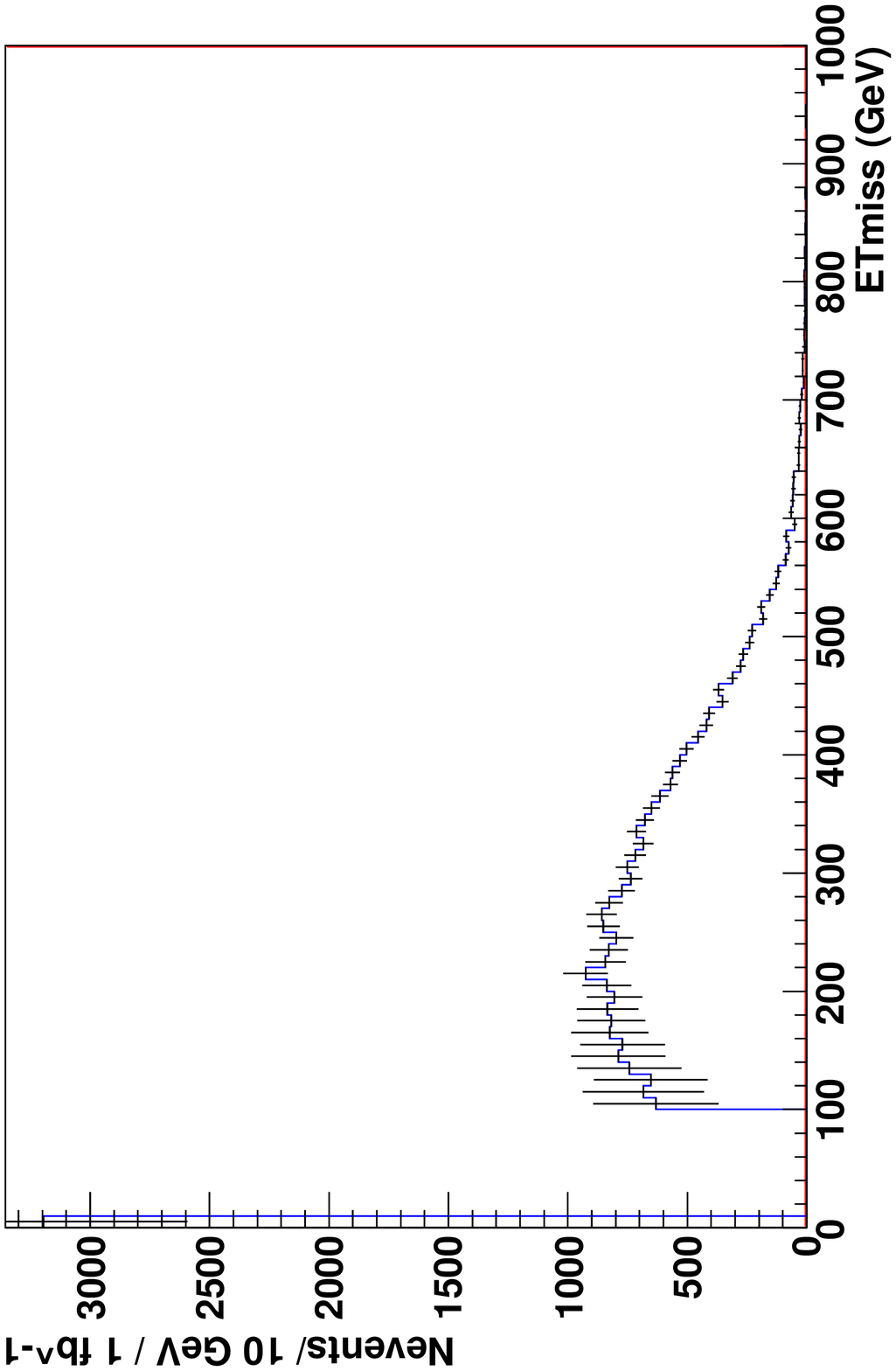}
%\put(130,50){$\Delta E_\tau$}
\put(90,50){$\scriptstyle  \slashed{E}_T^\text{(SM+BSM)-SM} $}
%\put(90,80){$\scriptstyle M_\text{inv}=0$}
\end{overpic}
\end{center}
\end{minipage}
\caption{\label{fig:diffM} Difference between total and SM
  $M_\text{vis}$ (Eq.~(\ref{eq:M})) (left) and $\slashed{E}_T$ (Eq.~(\ref{eq:Etmiss})) (right) distributions for the total inclusive sample. No further cuts were applied. }
\end{figure}

The strength of $\sqrt{\hat{s}}_\text{min}$ vs other (transverse) variables lies in the conjecture, given by Eqn. (\ref{eq:conj}), about the direct correlation between its peak position (given the correct mass $M_\text{inv}$) and the rise of the parton level production cross section. However, this correlation has so far not risen beyond the status of a conjecture. We therefore briefly discuss two other variables which might serve a similar purpose in background subtraction, namely 
\begin{\eqn}\label{eq:M}
M_\text{vis}\,=\,\sqrt{E^2_\text{vis}-\ora{P}^{2}_\text{vis}}
\end{\eqn} 
and 
\begin{\eqn}\label{eq:Etmiss}
\slashed{E}_{T}\,=\,|\slashed{\ora{P}}_T|,
\end{\eqn}
where everything is defined at analysis object level.
 In \cite{Konar:2008ei}, two more variables, namely $E_T$ and $H_T\,=\,E_T+\slashed{E}_{T}$,
 are studied\footnote{Note that the definition of $H_T$ differs in
 \cite{Konar:2010ma}, where it basically is set to the variable
 $M_\text{eff}$ \cite{Hinchliffe:1996iu,Tovey:2000wk}.}; however, as we define 
\begin{\eqn}\label{eq:Ptmiss}
\slashed{\ora{P}}_T\,=\,-\ora{P}_T
\end{\eqn}
these two variables are only variations of $\slashed{E}_T$ and therefore not discussed here. Figure \ref{fig:diffM} shows the subtracted $\text{(BSM+SM)}-\text{(SM)}$ distributions of $M_\text{vis}$ and $\slashed{E}_{T}$ respectively; especially the former looks quite promising. Indeed, a cut $M_\text{inv}\,>\,500\,\GeV$ reduces the SM background by a factor 6 ($W$ + jets) and 3 ($t\bar{t} $+jets), cf. Table \ref{tab:cuteff}. We equally compare to the frequently used variable $M_\text{eff}$ \cite{Hinchliffe:1996iu,Tovey:2000wk}
\begin{\eqn}\label{eq:meff}
M_\text{eff}\,=\,\sum_{\text{vis}}|p_T|\,+\,\slashed{E}_T,
\end{\eqn}
which exhibits a similar power for background suppression as $M_\text{vis}$, cf. Table \ref{tab:cuteff}.
 We therefore conclude that, for background suppression, both $M_\text{vis}, M_\text{eff}$ as well as $\sqrt{\hat{s}}_\text{min}$ work in a similar way; the advantage of the latter variable is the (conjectured) $M_\text{inv}$-dependent correlation between the its peak position and the hard (=parton level) process center-of-mass energy, if this can be proven to hold in all cases. Indeed, a similar conjecure of a linear correlation between the SUSY scale and the peak position of $M_\text{eff}$ \cite{Hinchliffe:1996iu,Tovey:2000wk} has recently been shown not to hold in all cases \cite{Conley:2010du}; however, an equivalent systematic study of $\sqrt{\hat{s}}_\text{min}$ is still lacking. % \footnote{The investigation of this correlation for several BSM scenarios on {\sl parton} level is currently under investigation \cite{ggowwork}.}.

\section{Conclusion and outlook}
We investigated the variable $\sqrt{\hat{s}}_\text{min}$
for a complete BSM sample which includes all strong production as well
as all possible decay chains. In our analysis, we include both soft as
well as detector effects by including a complete parton shower as well
as a generic detector simulation and reconstruction-level objects,
which have been defined such that the parton-level variable can be
recovered quite accurately. We investigate the variable  $\sqrt{\hat{s}}_\text{min}$ for a fully inclusive sample which sums over all possible final states of the hard scattering process, as well as for dominant exclusive final states. We see that, on parton level and for a correct input value of $M_\text{inv}$, the $\sqrt{\hat{s}}_\text{min}$ variable peaks closely to the heavy particle production cross section which is in agreement with the conjecture made by the authors of \cite{Konar:2008ei}. In a comparison between parton level and analysis level quantities, we see that in our sample the largest shift between these arises from the transition from tau-leptons which were used for the parton level quantity to the tau-jets and associated invisible neutrinos, which were used at analysis level. In order to asses this effect, we used the true tau-lepton four-vectors in the analysis level quantities. The effect is usually of the order of 100 \GeV, and in the totally inclusive sample the shift completely disappears for the idealized case of perfectly reconstructed tau jets. We therefore conclude that for the parameter point considered here, even at analysis level, the parton level $\sqrt{\hat{s}}_\text{min}$ peak position can be sufficiently reconstructed\footnote{We point out that, for the scenario considered in this study, even when using analysis tau jets the maximal shift between parton level and analysis level peak positions was $\sim\,200\,\GeV$, which effectively leads in an error $\sim\,100\,\GeV$ in the estimation of the initial heavy particles masses. The magnitude of this effect can of course differ depending on the BSM model and as well as specific model scenario point. }. In case the correlation between the threshold of the parton level cross section and the peak of the  $\sqrt{\hat{s}}_\text{min}$ distribution could be proven rigorously, this would indeed provide a quite elegant and straightforward way to assess the scale of the new physics signal as a function of the total invisible mass of the process.\\

Furthermore, we present the first study which investigates the use of
$\sqrt{\hat{s}}_\text{min}$ in order to suppress SM background for BSM
searches. For this, we considered $W$+ jets as well as $t\bar{t}$+jets
background. We saw that these backgrounds could be sufficiently
reduced by cuts on $\sqrt{\hat{s}}_\text{min}$, leading to suppression
factors around $2-6$, while we retained $90\%$ of the BSM signal. This
feature was independent of the input value of the total invisible mass
$M_\text{inv}$. However, we could achieve similar results by a cut on
the total visible mass $M_\text{vis}$, which is a simpler variable
which additionally does not require the input of $M_\text{inv}$. A
further comparison with $M_\text{eff}$ as a cut variable lead to
similar results. We therefore conclude that, unless the conjecture
about mass particle threshold and peak position of
$\sqrt{\hat{s}}_\text{min}$ can be rigorously proven, the latter does
not exhibit significant advantages over other (transverse)
variables. However, if the correlation between the threshold and
$\sqrt{\hat{s}}_\text{min}$ could be rigorously proven, it would
indeed provide a simple and elegant hold on the scale of new physics processes. A further investigation of this relation is in the line of future work.
\section*{Acknowledgements}
I want to thank the members of the 2009 Les Houches Mass determination group for providing me with the data as well as parts of the analysis framework used in this study, and specifically J.-R. Lessard and R. Bruneli\`ere for answering additional questions. Large parts of this work were inspired by discussions with members of the Cambridge SUSY group, and special thanks goes to B. Allanach, C. Lester, B. Webber, A. Papaefstathiou, and K. Sakurai for many valuable comments. I am equally grateful to K. Matchev for commenting on the manuscript prior to publication, as well as to P. Bechtle, M. Kobel, S. Wahrmund, and especially D. Kar for discussions of underlying event effects at a $14\,\TeV$ LHC. This work was supported by the STFC.
\begin{appendix}
\section*{Appendix}
\section{SPS1a spectrum}\label{sec:sps1aspectrum}
\begin{table}[h!]
\begin{\eqn*}
\begin{array}{|cc|cc|cc|cc|cc|cc|cc|cc|}\hline
\tilde{d}_{L}&568.4&\tilde{d}_{R}&545.2&\tilde{u}_{L}&561.1&\tilde{u}_{R}&549.3&
\tilde{b}_{1}&513.1&{\tilde{b}_{2}}&543.7&
{\tilde{t}_{1}}&399.7&{\tilde{t}_{2}}&585.8\\ %\hline
%&&&&&&&&&&&&&&&\\ %\hline
\tilde{l}_{L}&202.9&\tilde{l}_{R}&144.1&\tilde{\tau}_{1}&134.5&\tilde{\tau}_{2}&
206.9&\tilde{\nu}_{l}&185.3&\tilde{\nu}_{\tau}&184.7&\tilde{g}&607.7&& \\% \hline
%&&&&&&&&&&&&&&&\\ %\hline
\wt{\chi}^{-}_{1}&181.7&\wt{\chi}^{-}_{2}&380.0&\wt{\chi}^{0}_{1}&96.7&\wt{\chi}^{0}_{2}&181.1&|\wt{\chi}^{0}_{3}|&363.8&\wt{\chi}^{0}_{4}&381.7&&&&\\ \hline
\end{array}
\end{\eqn*}
\caption{Relevant masses for SPS1a in \GeV. $u\,=\,(u,c),\,d\,=\,(d,s),\,l\,=\,(e,\mu)$.}
\label{tab:sps1amasses}
\end{table}
\section{Minimization of $\sqrt{\hat{s}}$}\label{sec:mini}
As stated in Section \ref{sec:var}, $\sqrt{\hat{s}}_\text{min}$ denotes the {\sl minimal} center of mass energy $\sqrt{\hat{s}}$ of an event with a measured visible four-vector $P^\mu_\text{vis}\,=\,(E,\ora{P}_{T},P_{Z})$ which is still in agreement with total energy momentum conservation as well as onshellness of all outgoing particles. We equally assume the event to be at rest in the transverse plane such that Eqn. (\ref{eq:ptslash}) holds. The generic expression for $\hat{s}$ in this case is given by
\begin{\eqn}\label{eq:sstart}
\hat{s}_{(\ora{P}_{T}\,=\,-\ora{\slashed{P}}_{T})}\,=\,\lb E+\sum_j\,E_j \rb^2\,-\,\lb P_Z+\sum_j p_{j z} \rb^2
\end{\eqn}
where the index $j$ goes over the invisible particles in the event with the respective energies
\begin{\eqn*}
E^2_j\,=\,m_j^2+p_{j T}^2+p_{j z}^2.
\end{\eqn*}
Here and in the following, we omit the vector notation in $p_{T}$ for simplification, but all transverse quantities should be read as $p_{j T}\,=\,(p_{j x}, p_{j y}),\,\slashed{P}_T\,\,=\,(\slashed{P}_{X},\slashed{P}_{Y})$, etc.\\

\noindent
We use a Lagrange multiplier $\lambda$ to take the additional constraint for the vector sum of the transverse momenta
\begin{\eqn}\label{eq:cond}
\sum_j p_{j T}\,=\,\slashed{P}_{T}
\end{\eqn}
into account. We therefore aim at minimizing
\begin{\eqn*}
\mathcal{L}\,=\,\hat{s}-\lambda\lb \sum_j p_{j T}-\slashed{P}_{T}\rb,
\end{\eqn*}
ie we try to find the values of the invisible particles' three-momenta $\ora{p}_i$ such that
\begin{\eqn*}
\frac{\partial \mathcal{L}}{\partial \ora{p}_i}\,=\,0.
\end{\eqn*}
If we consider the case of $n_\text{inv}$ invisible particles in the event, we obtain $3\,\times\,n_\text{inv}$ equations
\begin{eqnarray}\label{eq:Leq}
\frac{\partial \mathcal{L}}{\partial p_{i T}}&=&2\,\lb E+\sum_j  E_j \rb\,\frac{p_{i T}}{E_i}-\lambda\,=\,0,\nonumber\\
\frac{\partial \mathcal{L}}{\partial p_{i z}}&=&2\,\lb E+\sum_j  E_j \rb\,\frac{p_{i z}}{E_i}\,-\,2\,\lb P_z\,+\,\sum_j p_{j z}\rb\,=\,0.
\end{eqnarray}
Together with the constraint in Eqn. (\ref{eq:cond}), we now have in total $3\,n_\text{inv}+1$ constraints for $3\,n_\text{inv}+1$ unknowns $\lb p_{i x}, p_{i y}, p_{i z};\lambda \rb$. From Eqns. (\ref{eq:Leq}), we immediately see that
\begin{\eqn*}
\frac{E_i}{E_j}\,=\,\frac{p_{i T}}{p_{j T}}\,=\,\frac{p_{i z}}{p_{j z}}\,=\,c_{ij}
\end{\eqn*}
with $c_{ij}\,=\,\frac{m_i}{m_j}\,\equiv\,\text{const}$. Combining this with Eqn. (\ref{eq:cond}), we obtain
\begin{\eqn}\label{eq:propconst}
\frac{m_i}{p_{i T}}\,=\,\frac{\sum_j m_j}{\slashed{P}_{T}}\,\equiv\,\frac{M_\text{inv}}{\slashed{P}_{T}},
\end{\eqn}
where we {\sl defined} $M_\text{inv}\,=\,\sum_j m_j$ to be the sum over the masses of all invisible particles, c.f. Eqn. (\ref{eq:minv}). We therefore have
\begin{\eqn*}
p_{iT}\,=\,\frac{\slashed{P}_{T}}{M_\text{inv}}\,m_i.
\end{\eqn*}
 %\end{appendix}
We can now rewrite the second equation in (\ref{eq:Leq}) and obtain
\begin{\eqn*}
\lb E+\frac{E_i}{p_{i T}}\slashed{P}_{T}\rb\frac{p_{i z}}{E_i}\,-\,\lb P_z+\frac{p_{i z}}{p_{i T}}\slashed{P}_{T}\rb\,=\,0.
\end{\eqn*}
Solving this for $p_{iz}$ leads to
\begin{\eqn*}
p_{i z}\,=\,\frac{P_z\,m_i}{\sqrt{E^2-P^2_z}}\,\sqrt{1+\frac{\slashed{P}^2_T}{M_\text{inv}^2}}.%,\,p_{T,i}\,=\,\frac{m_{i}\,\slashed{P}_{T}}{M_\text{inv}} 
\end{\eqn*}
We here have reproduced the solutions for $p_{i,z},\,p_{i,T}$ given in \cite{Konar:2008ei} which {\sl minimize} $\hat{s}$. Inserting these into Eqn.(\ref{eq:sstart}) then leads to $\sqrt{\hat{s}}_\text{min}$ given by Eqn. (\ref{eq:mastereq}), which denotes the {\sl minimal} hard scattering center of mass energy which is allowed by energy momentum conservation for a specific visible total four vector $(E,P_T,\,P_{Z})$ obtained from measurement. The only unknown quantity is $M_\text{inv}$ defined according to Eqn. (\ref{eq:minv}), which has to be treated as an external input parameter for $\sqrt{\hat{s}}_\text{min}$.\\

\noindent
We want to comment that the transverse mass variable $M_T$ \cite{vanNeerven:1982mz,Arnison:1983zy,Arnison:1985jk,Barger:1987du} has a functional form similar to $\sqrt{\hat{s}}_\text{min}$ as given in Eqn. (\ref{eq:mastereq}). For a system with visible and invisible total four-vectors $P^\mu_\text{vis},\,P^\mu_\text{inv}$, this variable is defined as
\begin{\eqn*}
M_T^2\,=\,\lb E_{T,\text{vis}}+E_{T,\text{inv}}\rb^2-\lb P_{T}+\slashed{P}_T\rb^2.
\end{\eqn*}
with the transverse energies
\begin{\eqn*}
E_{T,\text{vis}}^2\,=\,M_\text{vis}^2+P^2_T,\;E_{T,\text{inv}}^2\,=\,(M'_\text{inv})^2+\slashed{P}^2_T.
\end{\eqn*}
Here, $M_\text{vis},\, M'_\text{inv}$ denote the {\sl Lorentz-invariant masses} of the total visible and invisible system respectively,
\begin{\eqn*}
M_\text{vis}^2\,=\,P^{2}_\text{vis},\,
(M'_\text{inv})^2\,=\,P^{2}_\text{inv},
\end{\eqn*} 
which vary on an event by event basis.
Assuming Eqn. (\ref{eq:ptslash}) to hold, it follows that
\begin{\eqn*}
M_T\,=\,\sqrt{\slashed{E}_T^2+M_\text{vis}^2}\,+\,\sqrt{\slashed{E}_T^2+(M'_\text{inv})^2}.
\end{\eqn*}
We now see that the functional forms of $M_T(M'_\text{inv})$ and $\sqrt{\hat{s}}_\text{min} (M_\text{inv})$ as given in Eqn. (\ref{eq:mastereq}) are identical, and differences between the variables only stem from the difference between $M_\text{inv}'$ and $M_\text{inv}$.
Indeed, for a correct guess of $M_\text{inv}$, $\sqrt{\hat{s}}_\text{min}$ and $M_T$ coincide if for the invisible particles in the event
\begin{\eqn*}
\sum_{i\,\neq\,j} m_i m_j\,=\,\sum_{i\,\neq\,j} p_{i}\cdot p_{j}.
\end{\eqn*}
Componentwise, this equation is only fulfilled if we either have a complete set of massless particles which are all collinear with each other such that $\cos\theta_{ij}\,=\,1$ for all $(i,j)$ pairs or for a complete set of massive particles which are all produced at rest. In general, however,
\begin{\eqn*}
M'_\text{inv}\,>\,M^\text{(true)}_\text{inv}
\end{\eqn*} 
and therefore
\begin{\eqn*}
M_T\,>\,\sqrt{\hat{s}}_\text{min}\lb M_\text{inv}^\text{(true)}\rb
\end{\eqn*}
on an event by event basis.
\section{Pythia 6.4 ISR/ FSR default setup}\label{app:pshower}
All switch descriptions here are taken from \cite{Sjostrand:2006za}. We equally refer the reader to section 10 of this reference for a more detailed discussion of the parton shower model and its implementation in Pythia.
\vspace{3mm}\\

\begin{entry}
\iteme{MSTP(32) :} (D = 8) $Q^2$ definition in hard scattering for
$2 \to 2$ processes. For resonance production $Q^2$ is always chosen
to be $\hat{s} = m_R^2$, where $m_R$ is the mass of the resonance.\\
The newer options 6--10 are specifically intended for processes with 
incoming virtual photons. These are ordered from a `minimal'
dependence on the virtualities to a `maximal' one, based on
reasonable kinematics considerations. The old default value
\ttt{MSTP(32) = 2} forms the starting point, with no dependence at 
all, and the new default is some intermediate choice. 
Notation is that $P_1^2$ and $P_2^2$ are the virtualities of the 
two incoming particles, $\pT$ the transverse momentum of the 
scattering process, and $m_3$ and $m_4$ the masses of the two 
outgoing partons. For a direct photon, $P^2$ is the photon
virtuality and $x=1$. For a resolved photon, $P^2$ still refers
to the photon, rather than the unknown virtuality of the 
reacting parton in the photon, and $x$ is the momentum fraction
taken by this parton.
\begin{subentry}
\iteme{= 2 :} $Q^2 = (m_{\perp 3}^2 + m_{\perp 4}^2)/2 =
\pT^2 + (m_3^2 + m_4^2)/2$.
\iteme{= 8 :} $Q^2 = \pT^2 + (P_1^2 + P_2^2 + m_3^2 + m_4^2)/2$.  
ensure that the $Q^2$ scale is always bigger than $P^2$.  
\end{subentry}
\iteme{MSTP(62) :} (D = 3) level of coherence imposed on the
space-like parton-shower evolution.
\begin{subentry}
\iteme{= 3 :} $Q^2$/$\pT^2$ values and opening angles of emitted
(on-mass-shell or time-like) partons are both strictly ordered,
increasing towards the hard interaction.
\end{subentry}
 
\iteme{MSTP(63) :} (D = 2) structure of associated time-like
showers, i.e.\ showers initiated by emission off the incoming
space-like partons in \ttt{PYSSPA}.
\begin{subentry}
\iteme{= 2 :} a shower may evolve, with maximum allowed time-like
virtuality set by phase space or by \ttt{PARP(71)} times the $Q^2$
value of the space-like parton created in the same vertex, whichever
is the stronger constraint.
\end{subentry}
 
\iteme{MSTP(64) :} (D = 2) choice of $\alphas$ and $Q^2$ scale
in space-like parton showers in \ttt{PYSSPA}.
\begin{subentry}
\iteme{= 2 :} first-order running $\alphas$ with argument
\ttt{PARP(64)}$k_{\perp}^2 = $\ttt{PARP(64)}$(1-z)Q^2$.
\end{subentry}
 
\iteme{MSTP(65) :} (D = 1) treatment of soft-gluon emission in
space-like parton-shower evolution in \ttt{PYSSPA}.
\begin{subentry}
\iteme{= 1 :} soft-gluon emission is resummed and included
together with the hard radiation as an effective $z$ shift.
\end{subentry}

\iteme{MSTP(66) :} (D = 5) choice of lower cut-off for initial-state
QCD radiation in VMD or anomalous photoproduction events, and matching
to primordial $\kT$.
\begin{subentry}
\iteme{= 1 :} for anomalous photons, the lower $Q^2$ cut-off  is the 
larger of \ttt{PARP(62)}$^2$ and \ttt{VINT(283)} or\ttt{ VINT(284)},
where the latter is the virtuality scale for the 
$\gamma \to \q \qbar$ vertex on the appropriate side of 
the event. The \ttt{VINT} values are selected logarithmically
even between\ttt{ PARP(15)}$^2$ and the $Q^2$ scale of the
parton distributions of the hard process.    
\iteme{= 4 :} a stronger damping at large $\kT$, like 
$\d \kT^2/(\kT^2 + Q^2/4)^2$ with $k_0 < \kT < \pTmin(W^2)$. 
Apart from this, it works like \ttt{= 1}.
\iteme{= 5 :} a $\kT$ generated as in \ttt{= 4} is added vectorially 
with a standard Gaussian $\kT$ generated like for VMD states.      
Ensures that GVMD has typical $\kT$'s above those of VMD,
in spite of the large primordial $\kT$'s implied by hadronic
physics. (Probably attributable to a lack of soft QCD
radiation in parton showers.)
\end{subentry}
 
\iteme{MSTP(67) :} (D = 2) possibility to introduce colour coherence 
effects in the first branching of the backwards evolution of an 
initial-state shower in \ttt{PYSSPA}; mainly of relevance for QCD 
parton--parton scattering processes.
\begin{subentry}
\iteme{= 2 :} restrict the polar angle of a branching to be smaller 
than the scattering angle of the relevant colour flow.
\itemc{Note 1:} azimuthal anisotropies have not yet been included.
\itemc{Note 2:} for subsequent branchings, \ttt{MSTP(62) = 3} is
used to restrict the (polar) angular range of branchings.
\end{subentry}
 
\iteme{MSTP(68) :} (D = 3) choice of maximum virtuality scale and 
matrix-element matching scheme for initial-state radiation. To this
end, the basic scattering processes are classified as belonging
to one or several of the following categories (hard-coded for each 
process): 
\begin{subentry}
\iteme{= 0 :} maximum shower virtuality is the same as the $Q^2$ choice 
for the parton distributions, see \ttt{MSTP(32)}. (Except that the
multiplicative extra factor \ttt{PARP(34)} is absent and instead
\ttt{PARP(67)} can be used for this purpose.) No matrix-element 
correction. 
\iteme{= 3 :} as \ttt{= 0}, but ME corrections are applied where available.
\end{subentry}
 
\iteme{MSTP(69) :} (D = 0) possibility to change $Q^2$ scale for parton 
distributions from the \ttt{MSTP(32)} choice, especially for $\ee$.
\begin{subentry}
\iteme{= 0 :} use \ttt{MSTP(32)} scale.
\end{subentry}
 
\iteme{MSTP(72) :} (D = 1) maximum scale for radiation off FSR dipoles
stretched between ISR partons in the new $\pT$-ordered evolution in
\ttt{PYPTIS}.
\begin{subentry}
\iteme{= 1 :} the $\pTmax$ scale of FSR is set as the $\pT$ 
production scale of the respective radiating parton.
Dipoles stretched to remnants do not radiate.
\end{subentry}
\end{entry}
\vspace{1mm}
The additional switches/ variables appearing above are given by
\vspace{2mm}\\
\begin{entry}
\iteme{PARP(15) :} (D = 0.5 GeV) lower cut-off $p_0$ used to define
minimum transverse momentum in branchings $\gamma \to \q\qbar$ in
the anomalous event class of $\gamma\p$ interactions, i.e.\ sets the
dividing line between the VMD and GVMD event classes. 
\iteme{PARP(62) :} (D = 1.~GeV) effective cut-off $Q$ or
$k_{\perp}$ value (see \ttt{MSTP(64)}), below which space-like
parton showers are not evolved. Primarily intended for QCD showers 
in incoming hadrons, but also applied to $\q \to \q \gamma$ 
branchings.
\iteme{PARP(64) :} (D = 1.) in space-like parton-shower evolution
the squared transverse momentum evolution scale $k_{\perp}^2$ is
multiplied by \ttt{PARP(64)} for use as a scale in $\alphas$ and
parton distributions when \ttt{MSTP(64) = 2}.
\iteme{PARP(67) :} (D = 4.) the $Q^2$ scale of the hard scattering
(see \ttt{MSTP(32)}) is multiplied by \ttt{PARP(67)} to define the
maximum parton virtuality allowed in $Q^2$-ordered space-like 
showers. This does not apply to $s$-channel resonances, where the m
aximum virtuality is set by $m^2$. It does apply to all user-defined 
processes,however. 
\iteme{PARP(71) :} (D = 4.) the $Q^2$ scale of the hard scattering
(see \ttt{MSTP(32)}) is multiplied by \ttt{PARP(71)} to define the
maximum parton virtuality allowed in time-like showers. This does not
apply to $s$-channel resonances, where the maximum virtuality is set
by $m^2$. Like for \ttt{PARP(67)} this number is uncertain.
\iteme{VINT(283), VINT(284) :} virtuality scale at which a
GVMD/anomalous photon on the beam or target side of the event is 
being resolved. More precisely, it gives the $\kT^2$ of the 
$\gamma \to \q\qbar$ vertex. For elastic and diffractive scatterings, 
$m^2/4$ is stored, where $m$ is the mass of the state being diffracted.
For clarity, we point out that elastic and diffractive events are
characterized by the mass of the diffractive states but without
any primordial $\kT$, while jet production involves a primordial $\kT$
but no mass selection. Both are thus not used at the same time,
but for GVMD/anomalous photons, the standard (though approximate) 
identification $\kT^2 = m^2/4$ ensures agreement between the two
applications.
\end{entry}
\vspace{3mm}
\noindent
VDM/ GVDM are acronyms for vector meson dominated/ generalized vector meson dominated events in photo production respectively (cf section 7.7.2 of \cite{Sjostrand:2006za}).
\end{appendix}
%\newpage
\bibliography{../../lit}
\end{document}